\documentclass[pdflatex,sn-mathphys-ay]{sn-jnl}
\usepackage{longtable}
\usepackage{rotating}
\usepackage{amssymb}
\usepackage{multirow}
\usepackage{graphicx}
\usepackage{listings}
\usepackage{subcaption}
\usepackage{hyperref}
\usepackage[table]{xcolor}
\usepackage{hhline}
\usepackage{threeparttable}
\usepackage{array}
\usepackage[most]{tcolorbox}
\usepackage{hyphenat}
\usepackage{booktabs}

\usepackage{makecell}
\newcolumntype{Y}{>{\raggedright\arraybackslash}X}
\newcolumntype{C}[1]{>{\centering\arraybackslash}p{#1}}
\newcolumntype{L}[1]{>{\raggedright\arraybackslash}p{#1}}

\definecolor{lowgreen}{HTML}{D9EAD3}
\definecolor{medorange}{HTML}{FCE5CD}
\definecolor{highpurple}{HTML}{EADCF8}

\definecolor{unsatred}{HTML}{B00020}
\definecolor{satblue}{HTML}{1F4E79}
\definecolor{robustgreen}{HTML}{2E7D32}

\newcommand{\Low}{\colorbox{lowgreen}{\strut\textbf{Low}}}
\newcommand{\Med}{\colorbox{medorange}{\strut\textbf{Med}}}
\newcommand{\High}{\colorbox{highpurple}{\strut\textbf{High}}}

\newcommand{\Unsat}{\textcolor{unsatred}{\textbf{unsatisfied}}}
\newcommand{\Sat}{\textcolor{satblue}{\textbf{satisfied}}}
\newcommand{\Saty}{\textcolor{satblue}{\textbf{satisfy}}}
\newcommand{\RobustSat}{\textcolor{robustgreen}{\textbf{robustly satisfy}}}

\newcommand{\base}{ABLATION}
\newcommand{\app}{MELA}

\newcommand\inputsignal{\ensuremath{u}}
\newcommand\inputs{\ensuremath{\overline{\texttt{u}}}}

\newcommand\outputs{\ensuremath{\overline{\texttt{o}}}}

\newcommand\outputsignal{\ensuremath{o}}

\definecolor{kwcolor}{RGB}{0,102,204}
\definecolor{idcolor}{RGB}{20,20,20}
\definecolor{cmtcolor}{RGB}{120,120,120} 
\definecolor{lncolor}{RGB}{150,150,150}    

\lstdefinelanguage{NuSMVCustom}{
  morekeywords={
    MODULE,IVAR,VAR,DEFINE,ASSIGN,TRANS,SPEC,LTLSPEC,
    case,esac,init,next,TRUE,FALSE
  },
  sensitive=true,
  morecomment=[l]{--},
}

\lstdefinestyle{nusmvstyle}{
  language=NuSMVCustom,
  backgroundcolor=\color{white},
  basicstyle=\scriptsize\ttfamily\color{idcolor},
  keywordstyle=\color{kwcolor}\bfseries,
  commentstyle=\color{cmtcolor}\itshape,
  identifierstyle=\color{idcolor},
  numbers=left,
  numberstyle=\scriptsize\color{lncolor},
  numbersep=10pt,
  xleftmargin=2em,
  frame=single,
  framerule=0.3pt,
  rulecolor=\color{black!30},
  showstringspaces=false,
  keepspaces=true,
  columns=fullflexible,
  tabsize=2,
  breaklines=true
}

\begin{document}

\title[Synthesizing Behavioural Models of CPS Using Automata Learning and Statistical Machine Learning]{Synthesizing Behavioural Models of Cyber-Physical Systems Using Automata Learning and Statistical Machine Learning}

\author[1]{\fnm{Negin} \sur{Ayoughi}}\email{negin.ayoughi@uottawa.ca}

\author[1]{\fnm{Baharin A.} \sur{Jodat}}\email{balia034@uottawa.ca}

\author[1]{\fnm{Armina} \sur{Faghihi}}\email{afagh007@uottawa.ca}

\author[2]{\fnm{Patricio} \sur{Saavedra}}\email{pat@rabbit.run}

\author*[1]{\fnm{Shiva} \sur{Nejati}}\email{snejati@uottawa.ca}

\author[1]{\fnm{Mehrdad} \sur{Sabetzadeh}}\email{m.sabetzadeh@uottawa.ca}

\affil[1]{\orgdiv{School of Electrical Engineering and Computer Science},
\orgname{University of Ottawa},
\orgaddress{\city{Ottawa}, \state{Ontario}, \country{Canada}}}

\affil[2]{\orgname{RabbitRun Technologies Inc.},
\orgaddress{\state{Ontario}, \country{Canada}}}

\abstract{
Inferring behavioural models from system executions is essential for supporting formal verification and analysis of complex, heterogeneous cyber-physical systems (CPS).
Automata learning provides an effective way to infer state machine models from system executions. However, CPS inputs and outputs often consist of numeric time-series data, while automata learning algorithms  assume inputs over a finite symbolic alphabet. As a result, raw numeric data must first be abstracted into a finite set of symbols. In this article, we present \app, a passive automata learning approach enhanced with machine learning to synthesize behavioural models from numeric time-series data generated by CPS. \app\ systematically combines statistical machine learning with automata learning to automatically abstract raw numeric signals into interpretable intervals that are strongly correlated with system states. Specifically, \app\ uses information-theoretic variable selection and decision-tree-based range abstraction to transform numeric traces into symbolic representations suitable for automata learning. We evaluate \app\ on two CPS: a commercial network intrusion detection system developed by our industry partner, RabbitRun Technologies, and a publicly available industrial autopilot benchmark from the aerospace domain. Compared with expertise-based numeric data abstraction, \app\ reduces the number of states and transitions in the learned state machines by $49.20\%$ on average, while improving accuracy by $41.71\%$ on average. Furthermore, the learned state machines support system-level requirement verification and help practitioners explore behaviours that are not explicit in the system requirements. We make our implementation and experimental data available online~\citep{MELARepo}.}

\keywords{Automata learning, Cyber-physical systems, Behavioural model synthesis, Decision trees, Model checking, Intrusion detection, Simulink.}

\maketitle

\section{Introduction}
\label{sec:intro}
Cyber-physical systems (CPS) are complex and heterogeneous systems in which software controllers interact with physical processes through different sensors, actuators, and communication mechanisms. Their behaviour results from the interactions of many components, and their executions often generate long traces. Since this behaviour is difficult to understand directly, inferring state machines from execution data has been proposed as a way to obtain a high-level, interpretable, yet formal representation of system behaviour~\citep{hranisavljevic2016novel, medhat2015framework}. Such state-machine models help engineers better understand system behaviour and enable formal verification and validation against requirements.

Automata learning has long been used to learn state machines that capture system behaviour~\citep{Vaandrager17,muskardin2022active,muvskardin2022aalpy,GarhewalD23,NeiderSVK97}. Automata learning aims to infer a finite-state behavioural model of a black-box reactive system from observed input/output behaviour. It has been successfully applied to a range of real-world systems, including communication and security protocols such as BLE~\citep{PferscherA22}, MQTT~\citep{tappler2017model}, and TCP~\citep{fiteruau2016combining}, as well as an elevator controller~\citep{ovsiannikova2018active}, a biometric passport~\citep{aarts2010inference}, and autonomous driving  systems~\citep{hajnorouzi2025model}. Automata learning can be performed in either active or passive mode. In active learning, the assumption is that a well-defined and controllable interface for the system under learning is available, and the learner interacts with the system by issuing queries and observing the resulting outputs; in passive learning, the learner infers a model from offline execution data such as logs or traces. Active learning requires repeated interaction with the system under learning, as well as a reliable mechanism to query the system and steer it into different states. As a result, active learning can be costly and difficult to deploy in CPS, especially when online interaction and repeated execution are time-consuming and expensive, when a well-defined system interface is not available, and when  conditions for steering the system into different states are unknown. For such systems, passive learning is considered a more efficient and effective alternative to active learning, provided that a diverse  dataset of system behaviours is available~\citep{muskardin2022active}.

Regardless of these differences, both active and passive automata learning face a fundamental challenge when applied to CPS: \emph{their inputs and outputs are often continuous, numeric signals that evolve over time rather than symbols from a small finite alphabet.} Since automata-learning algorithms are inherently symbolic, they cannot operate directly on raw numeric time-series data and therefore require these signals to be abstracted into a finite set of symbols before learning can be applied. In many existing automata-learning applications, this abstraction step is either assumed to be available a priori or developed manually for the system under learning, rather than being derived through a systematic and interpretable procedure. Recently, some approaches have sought to automate this abstraction step, for example by using clustering or neural-network-based techniques~\citep{medhat2015framework, TapplerMAK24, PlambeckBHF24,hranisavljevic2020discretization, hranisavljevic2016novel}. 
 For CPS, however, it is not sufficient to obtain discrete symbols; the abstractions must also reflect distinctions in system behaviour and remain interpretable to engineers. Hence, abstractions must be derived from the observed behaviour of the system, rather than from clustering that groups values by numeric similarity, and must be expressed in a form that engineers can validate and interpret.

We propose the \emph{MachinE Learning-enhanced passive Automata learning approach (\app)} to derive state machines for CPS with numeric, time-series inputs and outputs. \app\ exercises the system under learning (SUL) across diverse operating conditions using coverage-guided input generation~\citep{offuttTesting} and collects the resulting system execution traces. It then converts these  traces into symbolic traces suitable for automata learning by abstracting raw numeric values into a finite set of intervals. To make these intervals behaviourally meaningful, \app\ uses decision-tree learning to identify abstractions that are strongly associated with the system states. \app\ then applies passive automata learning~\citep{cano2010inferring} to the symbolic traces to infer state machines that capture SUL's behaviours. Finally, \app\ evaluates whether each learned state machine conforms to the observed behaviour of the SUL by measuring its accuracy on unseen executions. If the measured conformance falls below a user-defined threshold, \app\ generates additional executions and repeats the abstraction and learning process. Once the learned state machine satisfies the required conformance threshold, \app\ uses temporal-logic model checking~\citep{CimattiCGGPRST02} to verify the system requirements over the learned state machine.  

\textbf{Contributions.}
This article presents \app, a systematic approach for synthesizing interpretable behavioural models of CPS from numeric time-series executions. The key characteristic of \app\ is that it derives interval-based abstractions according to their ability to distinguish system states. This yields abstractions that are both interpretable to engineers and effective for passive automata learning. This article extends our previous conference paper~\citep{neginconf}, published at the 27th International Conference on Model Driven Engineering Languages and Systems (MODELS 2024). Compared to our earlier paper, this article offers major expansions in the following areas: 
\begin{itemize}
\item We generalize \app\ from an approach developed for a specific industrial network-intrusion-detection system into a general approach for synthesizing and analyzing behavioural models of CPS with numeric time-series inputs and outputs. We strengthen the model-learning process to automatically resolve potential inconsistencies introduced during abstraction and extend our previous work beyond behavioural-model synthesis to evaluating model accuracy and supporting the formal verification of system requirements.
   
\item We extend the empirical evaluation by applying \app\ to a second CPS case study from the aerospace domain, in addition to the network system presented in our previous work.

\item We demonstrate how \app-generated state machines identify unspecified behaviours and characterize the conditions under which requirements hold.
\end{itemize}

\textbf{Findings.}  To assess the benefits of \app's automated abstraction, we compare the state machines learned by \app\ with those learned by a variant of \app\ in which the automated abstraction step is replaced with expertise-based abstractions. We evaluate both approaches using two criteria: model complexity, measured by the number of states and transitions, and accuracy. Our results show that \app\ generates state machines with $49.20\%$ fewer states and transitions on average, while achieving $41.71\%$ higher accuracy on average than the expertise-based variant. These results show that combining automata learning with our automated abstraction leads to more concise state machines that more accurately represent the behaviour of the SUL. We further use the learned state machines to verify system requirements and show that the verification results align with domain knowledge or system documentation. Finally,  we use the learned state machines to examine behaviours that are not explicitly specified in the requirements. In the network case study, \app\ shows how intermediate states respond to different traffic conditions, revealing behaviour that the requirements leave unspecified. In the aerospace case study, \app\ identifies the input conditions under which the system satisfies altitude-reaching requirements.

\textbf{Replication Package.} Our framework, empirical data, and supplementary material are publicly available~\citep{MELARepo}.

\textbf{Organization.} Section~\ref{sec:motivation} motivates the need for synthesizing behavioural models for CPS. Section~\ref{sec:mela} presents \app, our approach for deriving state-machine models for CPS. Section~\ref{sec:eval} presents the evaluation.  Section~\ref{sec:relwork} compares our work with related work. 
Section~\ref{sec:con} concludes the article.

 \section{Motivation}
\label{sec:motivation}
\begin{figure}[t]
    \centering
    \includegraphics[width=0.85\linewidth]{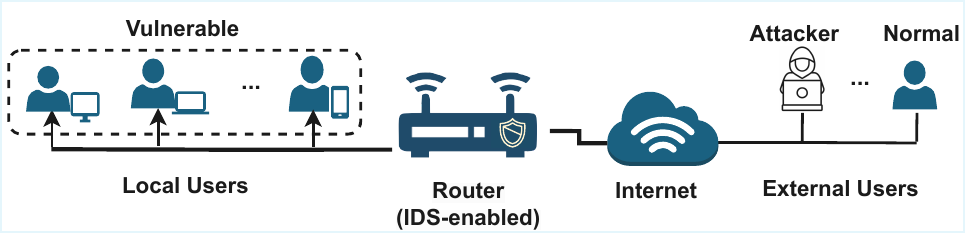}
   \caption{Real-world deployment of a router system enabled by a network intrusion detection system (IDS)}
   \label{fig:fig_1}
    
\end{figure}

We motivate the challenges of learning symbolic and interpretable behavioural models for numeric CPS using a real-world network intrusion detection system (IDS). Figure~\ref{fig:fig_1} provides an overview of a router system equipped with an IDS. The router connects local users to the Internet and external networks. It hosts an IDS that monitors external network traffic to identify denial-of-service (DoS) and distributed denial-of-service (DDoS) attacks originating from external users acting as attackers and targeting local users. DoS attacks flood their target with excessive traffic from a single source, whereas DDoS attacks employ multiple sources to perform a more extensive attack~\citep{zargar2013survey}. The IDS continuously monitors network traffic and updates its state based on features extracted from traffic flows. The IDS produces an output state indicating its assessment of the monitored traffic. This output state can be one of the following: \texttt{Safe}, indicating no signs of an attack; \texttt{Warning}, indicating unusual but not necessarily harmful traffic; \texttt{Tending Warning}, indicating elevated unusual traffic that may degrade network performance; \texttt{Tending Alert}, indicating highly suspicious traffic; and \texttt{Alert}, indicating harmful or malicious activity.

The IDS is expected to satisfy the following two requirements:

$\varphi_1 = $\textit{``When attacks happen, the system shall change state in a staged manner from safe to warning and from warning to alert''}, and 

$\varphi_2 = $\textit{``When attacks are stopped, the system shall restore in a staged manner its state from alert to warning, and from warning to safe.''}

These requirements  describe how IDS should move among the \texttt{Safe}, \texttt{Warning}, and \texttt{Alert} states when attacks start or stop. They are, however, only a partial specification of the IDS behaviour: while the IDS has five states -- \texttt{Safe}, \texttt{Warning}, \texttt{Alert}, \texttt{Tending Warning}, and \texttt{Tending Alert} -- the requirements refer only to the first three, leaving the behaviour of the two tending states unspecified. Engineers therefore need a behavioural model, learned from system executions, both to verify $\varphi_1$ and $\varphi_2$ and to examine how the IDS behaves in the \texttt{Tending Warning} and \texttt{Tending Alert} states under different traffic conditions.

Constructing a behavioural model for IDS raises a key challenge: the IDS operates on network traffic with numeric features. In particular, one important IDS input is the number of flows extracted from network traffic. To generate a behavioural model, this numeric input must be abstracted into symbolic values. Abstracting numeric values into symbols manually or in an ad-hoc manner can lead to poor behavioural models: an overly coarse abstraction may merge traffic conditions that correspond to different IDS behaviours, while an overly fine abstraction may introduce too many symbols and produce an unnecessarily complex model. \app\ therefore learns these abstractions automatically using decision-tree learning, which groups numeric values according to their association with the IDS output states. This yields interpretable symbols at an appropriate level of granularity: fine-grained enough to distinguish different IDS behaviours, yet coarse-grained enough to avoid unnecessarily complex models.

\begin{figure}[t]
    \centering
    \includegraphics[width=\linewidth]{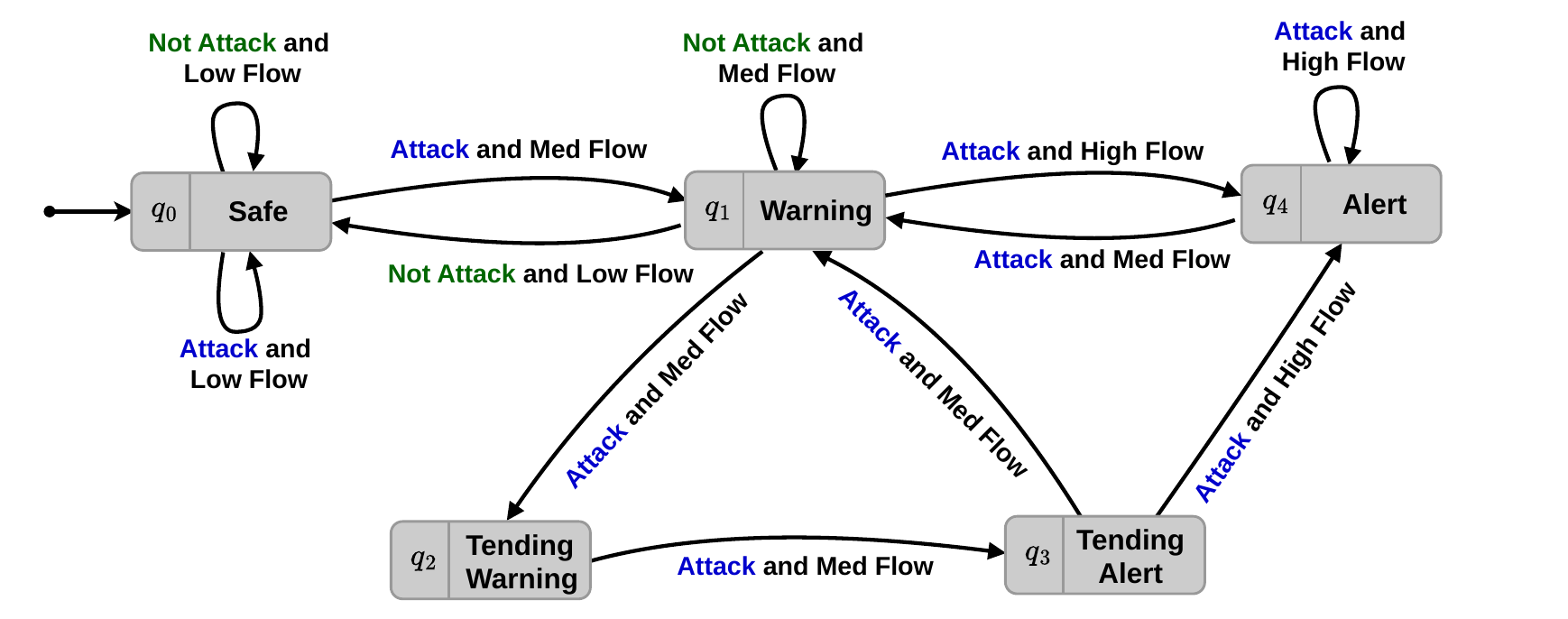}
    \caption{A \emph{simplified} example of a state machine learned for a network intrusion detection system (IDS) by our approach (\app).}
    \label{fig:SM}
       \vspace*{-.3cm}
\end{figure}

Figure~\ref{fig:SM} shows a simplified state machine learned by \app\ for the IDS. The model contains five states, corresponding to the five IDS operational states, and its transitions are labelled with abstract input symbols that characterize traffic conditions. These symbols capture two key aspects of the traffic: whether attack traffic is injected, represented by \texttt{Attack} and \texttt{Not\ Attack}, and the observed flow volume, represented by \texttt{Low\ Flow}, \texttt{Med\ Flow}, and \texttt{High\ Flow}. Rather than requiring engineers to define these flow-volume categories manually, \app\ derives them automatically using decision-tree learning. For the model in Figure~\ref{fig:SM}, \app\ partitions the numeric range of network flows, $[0,\infty)$, into three interpretable intervals: \texttt{Low\ Flow}=[0,454), \texttt{Med\ Flow}=[454,3500), and \texttt{High\ Flow}=$[3500,\infty)$. Thus, a transition labelled with \texttt{Attack} and \texttt{Med\ Flow} indicates that the IDS changes state when attack traffic is injected and the number of observed flows is between 454 and 3500. The state machine in Figure~\ref{fig:SM} supports two engineering tasks: verifying requirements $\varphi_1$ and $\varphi_2$, and exploring behaviour not specified by these requirements, namely the IDS behaviour in the \texttt{Tending Warning} and \texttt{Tending Alert} states.

 \section{ML-Enhanced Automata Learning (\app)}
\label{sec:mela}
\begin{figure}[t]
	\centering
        \includegraphics[width=\linewidth]{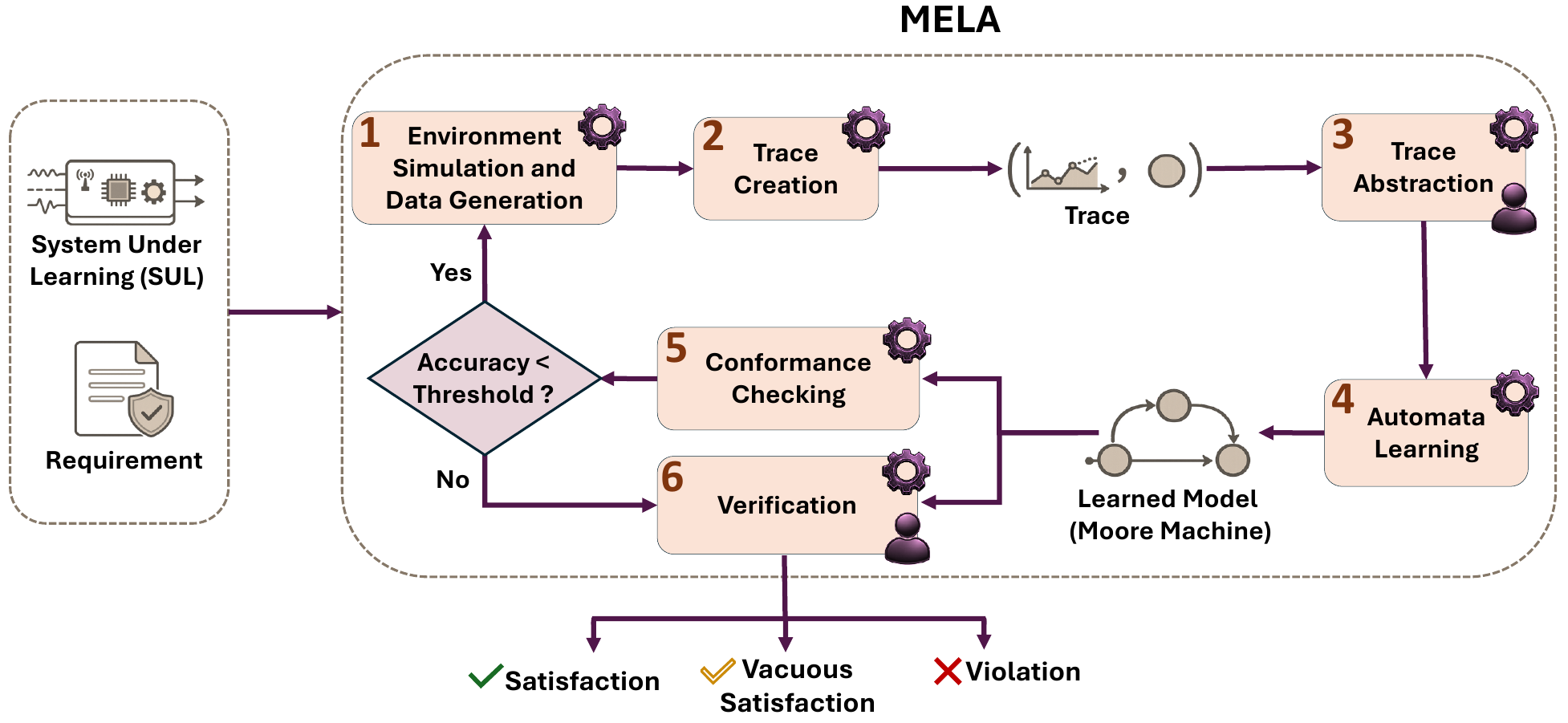}
\caption{Our approach for deriving and analyzing state machines, \app }
		\label{fig:mela}
     \vspace*{-.3cm}
\end{figure}

Figure~\ref{fig:mela} provides an overview of \app. \app\ takes as input a system under learning (SUL) and a set of requirements  for the SUL, and proceeds in six steps. First, \app\ executes the SUL to generate time-series data. Second, it converts the time-series data into traces suitable for automata learning. Third, \app\  selects inputs that are most predictive of the system state and therefore most relevant for specifying the learned state machine.
It then uses decision-tree learners to abstract the numeric value ranges of the selected inputs into symbolic categories. Fourth, \app\ learns state machines from the abstract traces produced in Step~3. Fifth, \app\ checks whether each learned state machine sufficiently conforms to the SUL by evaluating its accuracy against unseen system executions. If the accuracy falls below a user-defined threshold, \app\ returns to Step~1 to generate additional time-series data and repeats the learning process; otherwise, the state machine is passed to the final step. Sixth, \app\  verifies the requirements specified in temporal logic over the learned state machine and returns one of three outcomes: \emph{satisfaction}, \emph{vacuous satisfaction}, or \emph{violation}, depending on whether each requirement is satisfied, vacuously satisfied, or violated.

Below, we first provide the background needed for \app\ in Section~\ref{sec:background}. In Section~\ref{sec:example}, we describe the IDS discussed in Section~\ref{sec:motivation} in more detail to provide a running example that illustrates the different steps of \app. Next, in Sections~\ref{sec:step1} to~\ref{sec:step6}, we describe each step of \app\ in detail. Finally, in Section~\ref{sec:formalanalysis}, we discuss the determinism of the learned state machines and the soundness of the verification results obtained by \app.

\subsection{Background}
\label{sec:background}
In this section, we provide background on the formal notation for time-series data, the automata-learning algorithm used by \app, and the learned state-machine representation.

\textbf{Time-series data notation.} Provided with a SUL \texttt{S}, we denote an input for \texttt{S} as  $\inputs=(\inputsignal_1, \inputsignal_2 \ldots \inputsignal_m)$ and an output for \texttt{S}  as $\outputs= (\outputsignal_1, \outputsignal_2 \ldots \outputsignal_n)$ where $m$ is the number of system inputs, $n$ is the number of system outputs, and each $\inputsignal_j$ and each  $\outputsignal_j$  is a time-series vector for some input and some output of \texttt{S}, respectively. A time series is a function $v: [0, T] \rightarrow D$ where the interval $[0, T]$ is a time domain with duration $T$, and $D$ is the range of $v$. The time series $v$ can be continuous-valued when $D$ is an interval of real numbers (e.g., $[0,1]$). The time series $v$ can be discrete-valued when $D$ is a finite or countable set (e.g., $D = \{\texttt{UDP}, \texttt{TCP}\}$ or $D = \{1, \ldots, 6000\}$). As discussed in Sections~\ref{sec:intro} and~\ref{sec:motivation}, our goal is to abstract numeric input values, whether continuous or discrete, into symbolic categories that can be used to learn automata.

\textbf{Automata learning and state machine formalism.} \app\ learns a state machine using Regular Positive and Negative Inference (RPNI)~\citep{de2010grammatical,cano2010inferring,muskardin2022active,BergA25}, a well-established passive automata-learning algorithm. RPNI was originally developed for learning deterministic finite automata (DFAs), but its variants support richer automata models, including Mealy and Moore machines~\citep{muvskardin2022aalpy}. While Mealy machines associate outputs with transitions, Moore machines associate outputs with states. 

In our case studies, \app\ learns Moore machines because the observed outputs correspond to operational system states rather than transition-specific outputs. 
A Moore machine is a tuple \hbox{$\mathcal{M} = (I, O, Q, q_0, \delta, \lambda)$}, where $I$ is the set of input symbols, $O$ is the set of output symbols, $Q$ is the set of states, $q_0 \in Q$ is the initial state, $\delta: Q \times I \rightarrow Q$ is the transition function, and $\lambda: Q \rightarrow O$ is the state-labelling function, mapping each state to its output symbol~\citep{GiantamidisTB21}.

RPNI first constructs a prefix tree automaton (PTA) from the abstract traces~\citep{BergA25}. The PTA is a tree-shaped automaton in which each trace corresponds to a unique path from the initial state. Hence, before any generalization, the PTA captures exactly the behaviours observed in the traces. RPNI then generalizes the PTA by iteratively merging compatible states using the standard red/blue state-merging strategy~\citep{BergA25}. Starting from the root state, the algorithm repeatedly considers merges between red and blue states that have the same output symbol. Accepted merges generalize the observed behaviours and gradually produce a compact Moore machine.

\subsection{Running Example}
\label{sec:example}
\begin{figure}[t]
	\centering
        \includegraphics[width=\linewidth]{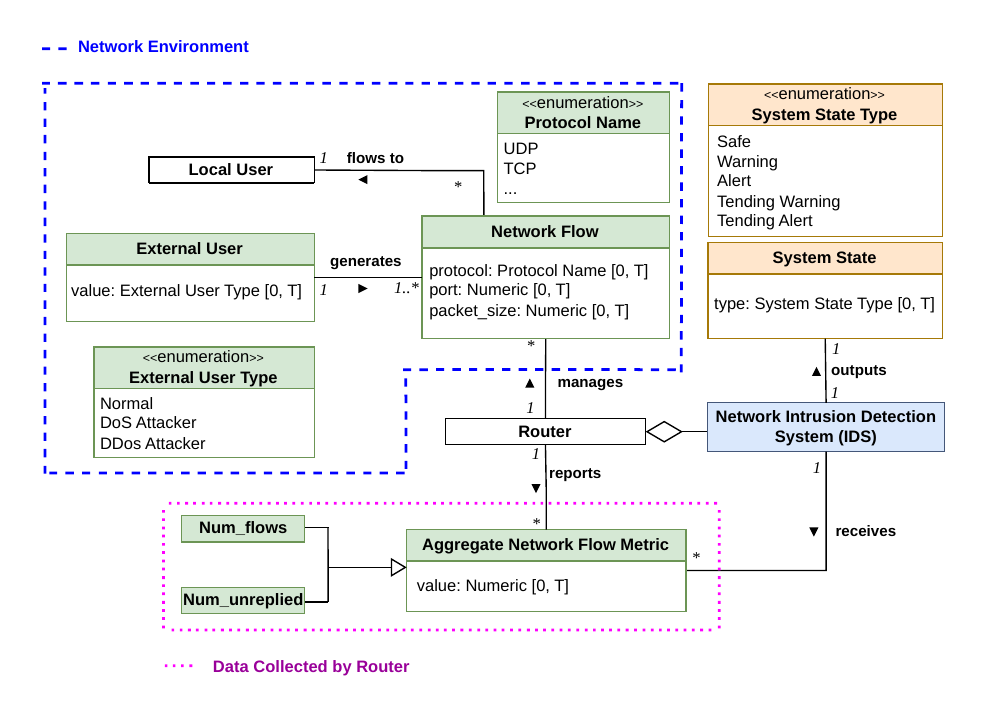}
\caption{Conceptual model for the IDS discussed in Section~\ref{sec:motivation}. The inputs are shaded green, the SUL is shaded blue, and the observable system-state output is shaded orange. The boundaries distinguish the network environment and the data collected by router.}
		\label{fig:conceptual}
\end{figure}

Figure~\ref{fig:conceptual} presents a conceptual model of the IDS introduced in Section~\ref{sec:motivation}, capturing its context, inputs, and outputs, so that the IDS can be used as a running example to illustrate the different steps of \app.
In Figure~\ref{fig:conceptual}, inputs are shaded green, the SUL blue, and the system-state output orange. The blue dashed boundary denotes the network environment, including external and local users and network flows, while the purple dotted boundary denotes router-extracted traffic features provided as IDS inputs.
As shown in the network environment, external users are either normal users or attackers (DoS/DDoS). External users generate network flows that are forwarded to local users through the router. Each flow is characterized by the communication  \texttt{protocol}; the  destination  \texttt{port} number of the incoming traffic; and \texttt{packet\_size}. The attribute \texttt{protocol} takes values from a finite set (e.g., $\{\texttt{TCP}, \texttt{UDP}\}$); \texttt{port} takes values in $[0,65535]$; and \texttt{packet\_size} is measured in bytes.

The router collects information from all flows traversing it and computes the aggregate flow metrics \texttt{num\_flows} and \texttt{num\_unreplied}. The metric \texttt{num\_flows} denotes the total number of flows passing through the router, while \texttt{num\_unreplied} captures the number of flows that remain unacknowledged by local users. 
The aggregate metrics (i.e., \texttt{num\_flows} and \texttt{num\_unreplied}), together with the per-flow attributes (i.e., \texttt{protocol}, \texttt{port}, \texttt{packet\_size}), are provided to the IDS operating on the router to update its operational state. 
The IDS selects one of the states \texttt{Safe}, \texttt{Warning}, \texttt{Tending Warning}, \texttt{Tending Alert}, or \texttt{Alert}.

\subsection{Step 1: Environment Simulation and Data Generation} 
\label{sec:step1}
In the first step, \app\ uses a simulator to exercise the SUL under different input stimuli by modelling its environment and how input values evolve over time. \app\ randomly generates inputs using existing parameterized time-series data generation techniques~\citep{signals19,pareto18}, executes the SUL, and collects the resulting outputs.

To ensure that the learned automata effectively capture the behaviours of the SUL, we need to generate  inputs that exercise the SUL under varied scenarios and yield  outputs that adequately capture the SUL's behaviours. To increase the adequacy of the generated data,  \app\ employs established black-box test coverage criteria for software testing based on system-state coverage~\citep{offuttTesting}. Specifically, after each execution, \app\ updates the achieved state coverage; when coverage no longer increases, \app\ stops generating additional inputs and proceed to the next step. Coverage saturation happens either when the generated outputs cover all system states or when after several consecutive iterations, the outputs do not cover any new states, indicating that data generation is unlikely to yield further improvements in state coverage.

\subsection{Step 2: Trace Creation} 
\label{sec:step2}
The second step of \app\ converts the time-series data generated in Step~1 into traces by sampling each time-series vector at a fixed period $\delta$. This yields finite, discrete-time traces suitable for automata learning. Specifically, each time-series vector $v : [0, T] \rightarrow D$ is converted into a sequence of samples $v^0, v^1, \ldots, v^k$, where $v^0 = v(0)$, $v^1 = v(\delta)$, $v^2 = v(2\cdot\delta)$, $\ldots$, and $v^k = v(k\cdot\delta)$, with $k\cdot\delta = T$. Let $\inputs = (\inputsignal_1, \inputsignal_2 \ldots \inputsignal_m)$ be an input of the SUL, and $\outputsignal$ be the output vector related to the system state.  
A trace $\tau$ corresponding to $\inputs$ and $\outputsignal$ is represented as follows:
\[
\begin{aligned}
\tau &= ((\inputsignal_1^0,\inputsignal_2^0, \ldots, \inputsignal_m^0,\outputsignal^0), (\inputsignal_1^1,\inputsignal_2^1, \ldots, \inputsignal_m^1,\outputsignal^1), \ldots, (\inputsignal_1^k,\inputsignal_2^k, \ldots, \inputsignal_m^k,\outputsignal^k))
\end{aligned}
\]

where $k$ is the number of steps with time-step size $\delta$ in time domain $[0,T]$. For each $l \in \{0, \ldots, m\}$, $\inputsignal^j_l$ denotes the value of the input vector $\inputsignal_l$ at step $j \in \{0,\ldots,k\}$, and $\outputsignal^j$ denotes the value of the output vector $\outputsignal$ at step $j$. 
We denote by $\mathit{TR}$ the set of traces generated in Step~2. 

For example, Figure~\ref{fig:trace_abstraction}(a) shows small excerpts from traces generated for the IDS discussed in Section~\ref{sec:motivation}. Each tuple consists of sampled IDS input values followed by the corresponding IDS state. Specifically, the values inside the brackets are, in order, sampled from the IDS inputs \texttt{External user type}, \texttt{port}, \texttt{protocol}, \texttt{packet\_size}, \texttt{num\_flows}, and \texttt{num\_unreplied}. The value outside the brackets is the IDS state observed for those inputs.

\begin{figure}[t]
\noindent

{\scriptsize\bfseries (a) Before trace abstraction\par}
\vspace{2pt}
\begin{lstlisting}[language=Python]
([Normal, 80, TCP, 173, 82, 60], Safe), ([DoS, 80, TCP, 122, 767, 640], Warning), ([DoS, 80, TCP, 178, 4331, 1250], Alert) ...
\end{lstlisting}

\vspace{0.8em}

{\scriptsize\bfseries (b) After trace abstraction\par}
\vspace{2pt}
\begin{lstlisting}[language=Python]
([Normal, Low, Low], Safe), ([DoS, Med, Med], Warning), ([DoS, High, High], Alert) ...
\end{lstlisting}

\caption{Traces for the IDS: (a) an example of an actual trace and (b) the same trace after trace abstraction.}
\label{fig:trace_abstraction}
\end{figure}

\subsection{Step 3: Trace Abstraction} 
\label{sec:step3}
This step consists of three substeps. \emph{First,} \app\ selects the SUL inputs that should be included in the learned state machine to enable verification of the system requirements.  \emph{Second,} \app\ abstracts the raw numeric ranges of the selected SUL inputs into symbolic categories and replaces the numeric values in the set  of traces obtained in Step~2  with their corresponding abstract symbols. Since converting numeric traces into abstract traces may lead to inconsistencies, i.e., traces with identical abstract inputs may be associated with different output states, in the \emph{third} substep, \app\ resolves any inconsistencies in the abstract traces. Below, we describe these three substeps, which we refer to as \emph{input  selection}, \emph{range abstraction}, and \emph{inconsistency resolution}, respectively.

\subsubsection{Input Selection}
\label{sec:varsel}
\app\ first prompts the user to identify the inputs referenced in the SUL requirements, since these inputs are relevant to the requirements and must be preserved for verification. \app\ then analyzes the remaining inputs by computing their information gain with respect to the system state, selecting those that are most predictive of the system state. This ensures that the learned state machine includes both the requirement-referenced inputs and the additional inputs needed to distinguish system states and characterize state transitions.

To this end, \app\ constructs a table from the set $\mathit{TR}$ of traces obtained from Step~2 by treating each sampled input-output tuple $(\inputsignal_1^j,\inputsignal_2^j,\ldots,\inputsignal_m^j,\outputsignal^j)$ from every trace $\tau_j \in\mathit{TR}$ as one row. The table contains one column for each input $\inputsignal_1,\ldots,\inputsignal_m$ and a final column for the system-state output $\outputsignal$. Using this table, \app\ computes the information gain of each input not referenced in the requirements, ranks these inputs by their information-gain values, and extends the set of requirement-referenced inputs with those whose information gain exceeds a threshold. Following prior work, \app\ sets this threshold based on the standard deviation of the information-gain values across all inputs~\citep{prasetiyowati2021determining}. Finally, \app\ refines the traces in $\mathit{TR}$ by removing all inputs that were not selected.

For example, for the IDS case study discussed in Section~\ref{sec:motivation}, requirements $\varphi_1$ and $\varphi_2$ describe the expected state changes when attack traffic starts and stops. Based on domain expertise, \texttt{External User type} is retained as a requirement-referenced input because it distinguishes normal users from attackers. \app\ then computes information gain over the remaining IDS inputs and selects \texttt{num\_flows} and \texttt{num\_unreplied} as the most predictive inputs for the IDS state. The final selected input set therefore consists of \texttt{External User type}, \texttt{num\_flows}, and \texttt{num\_unreplied}, and  other IDS inputs are removed from the traces.

\subsubsection{Range Abstraction} After input selection, \app\ derives interval-based abstractions for the selected numeric inputs using decision-tree learners. For each selected numeric input $u$, \app\ constructs a two-column table from the trace set $\mathit{TR}$. The first column contains sampled values of $u$, and the second column contains the corresponding system-state values. Specifically, for every trace $\tau \in \mathit{TR}$ and every sampling step $j$, \app\ adds one row of the form $(u^j,\outputsignal^j)$, where $u^j$ is the value of $u$ at step $j$ and $\outputsignal^j$ is the system state at that step. \app\ then trains a decision tree using the values of $u$ as inputs and the corresponding system-state values as categorical class labels. Each tree leaf records: (1) the number of samples assigned to the leaf (\emph{support}), and (2) the purity of the leaf (\emph{confidence}), indicating the degree to which the samples belong to the same system state. A higher purity therefore indicates a stronger association with a single system state. Each leaf is connected to its parent node through a condition of the form $u^j < c$, where $c$ is a constant within the range of $u^j$. \app\ selects the leaves whose support and purity both exceed user-defined thresholds and extracts the corresponding conditions linking those leaves to their parent nodes. Let $c_1, \ldots, c_r$ denote the constants appearing in the selected conditions in ascending order. \app\ uses these constants to partition the range of $u$ into the intervals $[0,c_1)$, $[c_1,c_2)$, $\ldots$, $[c_r,\infty)$. \app\ then assigns a unique abstract symbol to each interval and replaces every numeric sample $u^j$ in the traces of $\mathit{TR}$ with the symbol corresponding to the interval containing $u^j$.

\begin{figure}[t]
    \centering
        \includegraphics[width=0.83\linewidth]{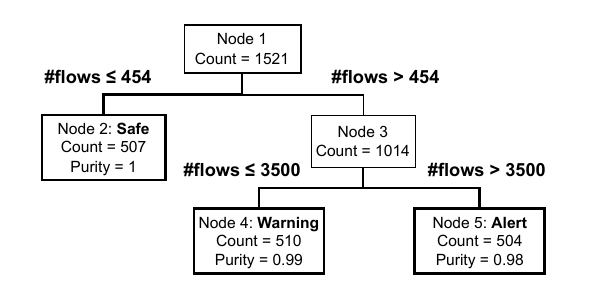}
    \vspace*{-.1cm}
    \caption{Illustrating how a decision tree is used to abstract the numeric range of the \texttt{num\_flows} input attribute. The numeric range of \texttt{num\_flows} is abstracted to the enumerated range [Low, Med, High] such that \texttt{Low}=$[0,454)$, \texttt{Med}=$[454,3500)$, and \texttt{High}=$[3500, \infty)$. }
    \label{fig:DT}
    \vspace*{-.1cm}
\end{figure}

For example, Figure~\ref{fig:DT} shows the decision tree used to abstract the numeric input \texttt{num\_flows}. As illustrated in Figure~\ref{fig:trace_abstraction}(b), the IDS traces contain tuples relating \texttt{num\_flows} and other IDS inputs to system states.  \app\ constructs the decision tree in Figure~\ref{fig:DT} . In this example, the maximum tree depth is set to three, while the support and purity thresholds are set to $20\%$ of the total data and $70\%$, respectively. Figure~\ref{fig:DT} shows the resulting tree leaves together with their support and purity values. \app\ selects Node~2, Node~4, and Node~5 because all three satisfy the support and purity thresholds. Based on the conditions linking these nodes to their parent nodes, \app\ partitions the range of \texttt{num\_flows} into the intervals $[0,454)$, $[454,3500)$, and $[3500,\infty)$. The number of intervals is determined automatically by the decision tree and is therefore not fixed a priori. For readability, these intervals are labelled as \texttt{Low}, \texttt{Med}, and \texttt{High}, respectively; specifically, \texttt{Low}=[0,454), \texttt{Med}=[454,3500), and \texttt{High}=$[3500,\infty)$. After input selection and range abstraction, it converts the numeric trace in Figure~\ref{fig:trace_abstraction}(a) into the abstract trace shown in Figure~\ref{fig:trace_abstraction}(b).

\subsubsection{Inconsistency Resolution} Converting numeric traces into abstract traces may introduce inconsistencies when the abstraction is too coarse. An inconsistency occurs when two traces share the same abstract input history but reach different output states. For example, consider two traces $\tau_1$ and $\tau_2$ with length three for a system with output states \texttt{state0}, \texttt{state1}, \texttt{state2}, and \texttt{state3}, and one numeric input $x$:

{\small
\[
\tau_1 =
([35], \texttt{state0}),\
([47], \texttt{state1}),\
([66], \texttt{state2})
\]
\[
\tau_2 =
([35], \texttt{state0}),\
([52], \texttt{state3}),\
([77], \texttt{state2})
.\]
}
In $\tau_1$ and $\tau_2$, the values in brackets denote values of $x$. Suppose range abstraction maps the range of  $x$ to three symbols: \texttt{Low} = $[0,40)$, \texttt{Med} = $[40,60)$, and \texttt{High} = $[60,\infty)$. This yields the following abstract traces: 

\newpage

\[
\hat{\tau}_1 =
([\texttt{Low}], \texttt{state0}),\
([\texttt{Med}], \texttt{state1}),\
([\texttt{High}], \texttt{state2})
\]
\[
\hat{\tau}_2 =
([\texttt{Low}], \texttt{state0}),\
([\texttt{Med}], \texttt{state3}),\
([\texttt{High}], \texttt{state2})
.\]
The abstract traces $\hat{\tau}_1$ and $\hat{\tau}_2$ are inconsistent in the second step: The first steps of the two traces are identical. But in the second step after receiving identical input 
[\texttt{Med}], the traces reach different states, namely \texttt{state1} and \texttt{state3}. Such traces cannot be used directly to learn a deterministic state machine~\citep{linz2022introduction}.

We define inconsistent traces as follows. Let $\bar{\tau}_1$ and $\bar{\tau}_2$ be two abstract traces of length $k$ obtained after range abstraction. The traces are inconsistent at step $j \leq k$ iff they are identical in the first $j-1$ steps, but at $j$th step  they have the same abstract inputs and different outputs. A trace set obtained after range abstraction is inconsistent if it contains at least one pair of inconsistent traces.

To resolve inconsistencies, \app\ proceeds  by refining the numeric-range abstraction by iteratively increasing the decision-tree depth, up to a user-defined maximum,  and reconstructing the induced intervals. This addresses inconsistencies caused by overly coarse abstractions. In the example above, the inconsistency between $\hat{\tau}_1$ and $\hat{\tau}_2$ can be resolved by partitioning \texttt{Med}=$[40,60)$ into finer intervals such as $[40,48)$ and $[48,60)$. The finer partition separates values that were previously mapped to the same abstract symbol but led to different system states.

Not all inconsistencies can be resolved by refining the abstraction. In CPS, the SUL may exhibit unstable behaviour, producing different outputs for the same inputs due to environmental uncertainty or noise. Moreover, increasing the decision-tree depth beyond a user-defined limit may overfit the data and increase the complexity of the learned state machine. Therefore, if inconsistencies persist after refinement, \app\ removes traces iteratively. Trace removal is used only as a last resort after abstraction refinement reaches the maximum user-defined depth and is intended to address residual inconsistencies arising from unstable or noisy CPS behaviour.
For each trace, \app\ counts its pairwise inconsistencies with other traces and removes the trace with the largest count. Ties are broken randomly. This process continues until the abstract trace set is consistent. As reported in Section~\ref{sec:eval}, in our case studies, range-abstraction refinement resolved most inconsistencies, and only $0.1\%$ of traces had to be removed.

\subsection{Step 4: Automata Learning} 
\label{sec:step4}
In the fourth step, \app\ uses the RPNI algorithm (Section~\ref{sec:background}) to learn a Moore machine from the abstract traces generated in Step~3. The input and output alphabets of the Moore machine are defined as follows. Let $m'$ be the number of inputs retained after input selection, and let $A_1, A_2, \ldots, A_{m'}$ denote the finite sets of abstract symbols associated with these retained inputs after range abstraction in Step~3. These sets are finite because Step~3 replaces each numeric range with a finite set of abstract symbols. The input alphabet of the Moore machine is then defined as $I = A_1 \times A_2 \times \cdots \times A_{m'}$.
Hence, each element of $I$ represents one possible combination of abstract values for the retained inputs and is treated as a single input symbol of the Moore machine. The output alphabet $O$ is the finite set of observable system states.

For example, Figure~\ref{fig:SM} shows a Moore machine learned for the IDS case study. In this machine, 
the abstract symbols for the number of flows are \texttt{Low Flow}, \texttt{Med Flow}, and \texttt{High Flow}, whereas the abstract symbols for external user type are \texttt{Attack} and \texttt{Not Attack}. The set of input symbols $I$ for this machine is 
I = \{
(\texttt{Not Attack, Low Flow}), 
 (\texttt{Not Attack, Med Flow}), 
 (\texttt{Not Attack, High Flow}), 
 (\texttt{Attack, Low Flow}),
 (\texttt{Attack, Med Flow}),
 (\texttt{Attack, High Flow})
\}.
The set of output symbols is
$
O = \{\texttt{Safe}, \texttt{Warning}, \texttt{Tending Warning}, \texttt{Tending Alert}, \texttt{Alert}\},
$
which corresponds to the IDS states.

RPNI can learn state machines from both positive traces, which capture behaviours to include, and negative traces, which capture behaviours to exclude. In our setting, \app\ uses only positive traces, namely the abstract traces generated in Step~3 from observed SUL executions. We do not provide negative traces because the available requirements and documentation do not precisely characterize invalid behaviours. Nevertheless, RPNI remains suitable because it can learn effective state machines from positive traces alone~\citep{bartocci2020mining, bartocci2021mining, agostinelli2021discovering}.

As discussed in Section~\ref{sec:background}, state merging in RPNI may introduce behaviours that do not appear in the original traces. Such generalization can affect verification results. To validate these results against observed executions, as we discuss in Section~\ref{sec:step6}, \app\ uses the PTAs generated by RPNI to distinguish results supported by the original traces from those caused only by state merging.

In the rest of this article, depending on the context, we interchangeably refer to Moore machines as either state machines or automata.

\subsection{Step 5: Conformance Checking} 
\label{sec:step5}
The fifth step of \app\ assesses whether the state machine learned in Step~$4$  conforms to the SUL. Following established practice in automata learning~\citep{muskardin2022active}, \app\ evaluates conformance using a test set, i.e., a set of traces generated from the SUL but not used during learning. This check is important because, in passive learning, the quality of the learned automaton depends on the completeness and diversity of the traces used for inference; if the learning traces cover only part of the SUL behaviour, the learned automaton may deviate from the actual system.

To measure conformance, \app\ first constructs an independent set of \emph{test traces}. Specifically, \app\ randomly generates data using Step~1, converts this data into traces using Step~2, and abstracts these traces using the interval abstractions learned in Step~3. \app\ then computes conformance as the percentage of test traces accepted by the learned automaton. A test trace is accepted if the automaton contains a path from the initial state that follows the trace’s abstract input symbols and produces the same system states as those in the trace. If the resulting conformance exceeds a user-defined threshold the automaton is considered sufficiently conformant and is used for requirements analysis. Otherwise, \app\ generates additional data and repeats the trace abstraction, automata learning, and conformance-checking steps.

\subsection{Step 6: Verification} 
\label{sec:step6}
The final step of \app\ uses temporal-logic model checking to verify SUL requirements on the Moore machines learned in Step~4. \app\ first translates the learned Moore machine into the input language of \textsc{NuSMV}~\citep{CimattiCGGPRST02}. \app\ then prompts users to encode the SUL requirements as temporal-logic properties, which are checked by \textsc{NuSMV} on the NuSMV code of the Moore machine. Because passive automata learning generalizes from a finite set of observed traces, the learned Moore machine may include behaviours that were not observed in the SUL executions. Therefore, when satisfaction or violation of a requirement may be affected by this generalization, \app\ uses the PTA produced by RPNI, as discussed in Section~\ref{sec:background}, to interpret the verification result with respect to the behaviours  observed in the SUL execution traces.  Based on this analysis, \app\ classifies each requirement as \emph{satisfied}, \emph{vacuously satisfied}, or \emph{violated}. Below, we explain the \textsc{NuSMV} encoding of learned Moore machines and present \app’s procedure for interpreting \textsc{NuSMV} results as verdicts over the SUL execution traces.

\begin{figure}[t]
\centering
\begin{minipage}{0.75\linewidth}
\begin{lstlisting}[
  style=nusmvstyle,
  basicstyle=\scriptsize\ttfamily,
  commentstyle=\color{gray}\ttfamily,
  numbers=left,
  numberstyle=\tiny\color{gray},
  stepnumber=1,
  numbersep=8pt,
  frame=single,
  xleftmargin=0pt,
  framexleftmargin=20pt,
  mathescape=true
]
MODULE main

VAR
  input : {$i_1$, $i_2$, ..., $i_m$};    -- input alphabet $\color{gray}{I}$
  state : {$q_0$, $q_1$, ..., $q_n$};    -- state set $\color{gray}{Q}$

DEFINE
  label := case
    state = $q_0$ : $\lambda(q_0)$;
    state = $q_1$ : $\lambda(q_1)$;
    ...
    state = $q_n$ : $\lambda(q_n)$;
  esac;                         -- output label of the current state

ASSIGN
  init(state) := $q_0$;          -- initial state

TRANS
  (state = $q$ & input = $i$ & next(state) = $\delta(q,i)$) |
  ... ;                         -- one disjunct for each defined transition $\color{gray}{\delta(q,i)}$ of the Moore machine

-- User-provided temporal-logic formulas of the SUL requirements. 
SPEC $\varphi$;       -- if $\color{gray}{\varphi}$ is a CTL formula
LTLSPEC $\varphi$;    -- if $\color{gray}{\varphi}$ is an LTL formula

\end{lstlisting}
\end{minipage}
\caption{Template for encoding a Moore machine 
$\mathcal{M} = (I, O, Q, q_0, \delta, \lambda)$ in the input language of NuSMV.}
\label{fig:mooretosmv}
\end{figure} 
\subsubsection{Moore Machine to the Input Language of \textsc{NuSMV}} 
\label{sec:nusmvcode}
Figure~\ref{fig:mooretosmv} presents the template used by \app\ to translate a Moore machine $\mathcal{M} = (I, O, Q, q_0, \delta, \lambda)$ into the input language of \textsc{NuSMV}. The input alphabet $I$ is encoded as the \textsc{NuSMV} variable \texttt{input}, while the state set $Q$ is encoded as the variable \texttt{state}. The initial Moore-machine state $q_0$ is specified by assigning it to \texttt{init(state)}. The state-labelling function $\lambda$ is encoded in the \texttt{DEFINE} block, where \texttt{label} maps each state $q \in Q$ to its corresponding output label $\lambda(q)$. The transition function $\delta$ is encoded in the \texttt{TRANS} block as a Boolean constraint over the current state, the current input, and the next state. Specifically, for each defined transition $\delta(q,i)$, the generated \textsc{NuSMV} model contains a disjunct stating that, when \texttt{state} is $q$ and \texttt{input} is $i$, the next value of \texttt{state} must be $\delta(q,i)$. This encoding therefore preserves the state-based outputs and transition structure of the learned Moore machine. User-provided temporal-logic requirements are then added to the generated model. 
These requirements are expressed in Computation Tree Logic (CTL)~\citep{clarke1986automatic} or Linear Temporal Logic (LTL)~\citep{Pnueli77}, two standard temporal logics for specifying system requirements~\citep{mcbook}. CTL expresses branching-time properties over computation trees and allows quantification over possible execution paths~\citep{clarke1986automatic}, whereas LTL expresses linear-time properties over individual executions~\citep{Pnueli77}. In \textsc{NuSMV}, CTL properties are specified using \texttt{SPEC}, while LTL properties are specified using \texttt{LTLSPEC}.

\begin{figure}[t]
\centering
\begin{minipage}{\linewidth}
\begin{lstlisting}[
  style=nusmvstyle,
  basicstyle=\scriptsize\ttfamily,
  commentstyle=\color{gray}\ttfamily,
  numbers=left,
  numberstyle=\tiny\color{gray},
  stepnumber=1,
  numbersep=8pt,
  frame=single,
  xleftmargin=0pt,
  framexleftmargin=20pt,
  mathescape=true
]
MODULE main

VAR
  input : {notattack_low_flow, attack_low_flow, notattack_med_flow,
           attack_med_flow, notattack_high_flow, attack_high_flow};
  state : {$q_0$, $q_1$, $q_2$, $q_3$, $q_4$};

DEFINE
  label := case
    state = $q_0$ : Safe;
    state = $q_1$ : Warning;
    state = $q_2$ : TendingWarning;
    state = $q_3$ : TendingAlert;
    state = $q_4$ : Alert;
  esac;    

ASSIGN
  init(state) := $q_0$;

TRANS
  (state = $q_0$ & input = notattack_low_flow & next(state) = $q_0$) |
  (state = $q_0$ & input = attack_low_flow & next(state) = $q_0$) |
  (state = $q_0$ & input = attack_med_flow & next(state) = $q_1$) |

  (state = $q_1$ & input = notattack_low_flow & next(state) = $q_0$) |
  (state = $q_1$ & input = notattack_med_flow & next(state) = $q_1$) |
  (state = $q_1$ & input = attack_med_flow & next(state) = $q_2$) |
  (state = $q_1$ & input = attack_high_flow & next(state) = $q_4$) |

  (state = $q_2$ & input = attack_high_flow & next(state) = $q_3$) |

  (state = $q_3$ & input = attack_med_flow & next(state) = $q_1$) |
  (state = $q_3$ & input = attack_high_flow & next(state) = $q_4$) |

  (state = $q_4$ & input = attack_med_flow & next(state) = $q_1$) |
  (state = $q_4$ & input = attack_high_flow & next(state) = $q_4$);

-- User-provided temporal-logic specifications for the SUL requirements $\color{gray}{\varphi_1}$ and $\color{gray}{\varphi_2}$
SPEC AG ((input = attack_high_flow & label = Safe) -> AX (label = Warning | label = TendingWarning))
SPEC AG ((input = notattack_low_flow & label = Warning) -> AX (label = Safe))

-- Specifications added by MELA for vacuity checking
SPEC EF (input = attack_high_flow & label = Safe)
SPEC EF (input = notattack_low_flow & label = Warning)

\end{lstlisting}
\end{minipage}
\caption{NuSMV code for the Moore machine in Figure~\ref{fig:SM} using the template in Figure~\ref{fig:mooretosmv}.}
\label{fig:sm2-nusmv}
\end{figure} 
Figure~\ref{fig:sm2-nusmv} shows the \textsc{NuSMV} code capturing the Moore machine in Figure~\ref{fig:SM}. In this Figure, lines $39$ and $40$, show two user-provided CTL formulas that partially capture requirements $\varphi_1$ and $\varphi_2$ described in Section~\ref{sec:motivation}.  In CTL, the operator ``AG'' means that the formula must hold globally on all computation paths, and the operator ``AX'' means that the formula must hold in all immediate successor states. Specifically, line 39 states that whenever the IDS is in \texttt{Safe} and observes \texttt{attack\_high\_flow}, all next states must be either \texttt{Warning} or \texttt{TendingWarning}. This partially captures $\varphi_1$ by requiring a staged response to attacks. Line 40 states that whenever the IDS is in \texttt{Warning} and observes \texttt{notattack\_low\_flow}, all next states must be \texttt{Safe}. This partially captures $\varphi_2$ by requiring recovery toward the safe state when attacks stop. The full encoding of   $\varphi_1$ and $\varphi_2$ in CTL is provided in our supplementary material~\citep{MELARepoEvaluation}.

\subsubsection{Interpreting Verification Results}
\label{sec:reqver} 
\begin{figure}[t]
    \centering
        \includegraphics[width=\linewidth]{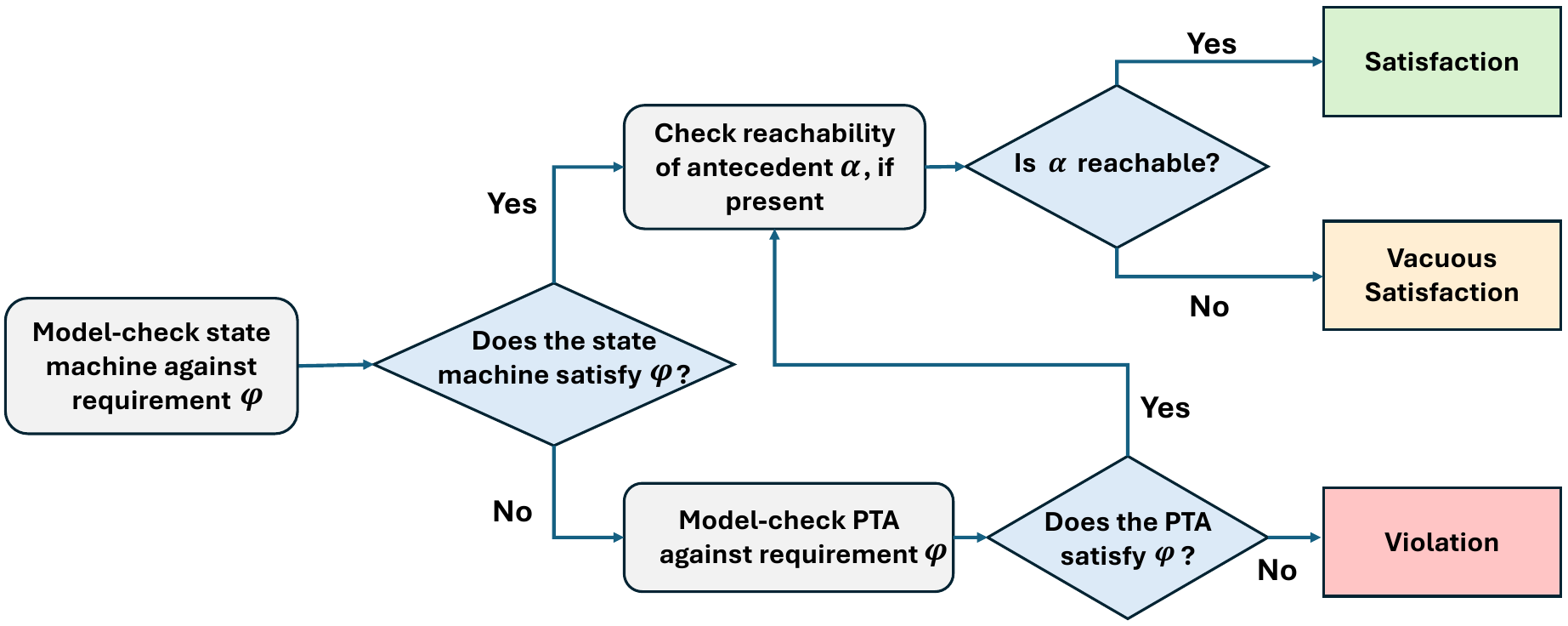}
\caption{\app's process in Step~$6$ for interpreting the verification results.}
    \label{fig:decision}
\end{figure}

Figure~\ref{fig:decision} shows how \app\ interprets the satisfaction and violation results returned by \textsc{NuSMV}. The interpretation differs for universal, existential, and hybrid properties. In this section, we focus on universal properties, e.g., the properties on lines 39 and 40 in Figure~\ref{fig:sm2-nusmv}, since all properties in our case studies are captured as universal properties. The interpretation process for existential and hybrid properties, which is similar to the one discussed here, is provided in the supplementary material~\citep{MELARepo}.

Based on Figure~\ref{fig:decision}, when \textsc{NuSMV} reports that a requirement is satisfied, \app\ checks whether the satisfaction is vacuous. Vacuous satisfaction can occur for implication requirements when the antecedent is never reached along any path~\citep{BeerBER97}. For a CTL requirement of the form $AG(\alpha \rightarrow \beta)$, \app\ checks the auxiliary formula $EF(\alpha)$, which holds when $\alpha$ is reachable along some path. If $EF(\alpha)$ is violated, then $\alpha$ is unreachable and the requirement is satisfied vacuously; otherwise, the antecedent occurs in at least one reachable state, and the satisfaction is non-vacuous. Lines~$43$ and~$44$ in Figure~\ref{fig:sm2-nusmv} show examples of the additional CTL specifications generated by \app\ for this purpose.
For an LTL requirement of the form $G(\alpha \rightarrow \beta)$, \app\ checks the auxiliary formula $G(\neg \alpha)$. If $G(\neg \alpha)$ is satisfied, then $\alpha$ never occurs on any path, and the requirement is satisfied vacuously. Otherwise, if $G(\neg \alpha)$ is violated, then the requirement is satisfied non-vacuously. 

When \textsc{NuSMV} reports that a requirement is violated, \app\ checks whether the violation is also exhibited by the PTA. To do so, \app\ encodes the PTA in \textsc{NuSMV} using the template in Figure~\ref{fig:mooretosmv}, since the PTA is a tree-structured Moore machine whose states, transitions, and output labels can be directly represented in \textsc{NuSMV}. Because \textsc{NuSMV} evaluates CTL requirements over infinite paths, \app\ adds a self-loop to the last state of each PTA trace to encode finite traces as infinite stuttering executions. Since the PTA captures behaviours before state merging, this check distinguishes violations supported by observed traces from those introduced by state merging. If the PTA also violates the requirement, the violating behaviour is supported by the observed traces, and \app\ reports the requirement as violated. If the PTA satisfies the requirement, the violation in the learned Moore machine is attributed to state merging. In this case, \app\ reports vacuous satisfaction if the requirement is satisfied vacuously on the PTA; otherwise, it reports satisfaction.

\subsection{Determinism, Soundness, and Vacuity} 
\label{sec:formalanalysis}
In this section, we discuss two formal properties of the state machines produced by \app: determinism and soundness. We also explain how vacuous satisfaction verdicts should be interpreted when reasoning about the SUL.

\textbf{Determinism.} A state machine is deterministic if each state-input pair has at most one successor state. The state machines learned by \app\ are deterministic by construction. Specifically, Step~3 resolves inconsistencies introduced by abstraction, ensuring that the abstract trace set used for learning is consistent. RPNI then learns a deterministic Moore machine~\citep{oncina1992identifying}.

\textbf{Soundness.} In our setting, soundness means that a satisfaction or violation verdict reported by \app\ is (1)~guaranteed with respect to the observed SUL executions used for learning, and (2)~is empirically supported with respect to broader SUL behaviour through  conformance checking:  First, Step~6 provides an analytical guarantee with respect to the learning traces: \app\ reports a requirement as satisfied or violated only when the verdict holds on the learned state machine and, when state merging may affect the outcome, the verdict is checked against the PTA constructed directly from SUL executions. Thus, the reported verdict is guaranteed to hold for the behaviours represented in the learning traces. Second, Step~5 provides empirical support that these traces are representative of the SUL behaviour by evaluating the learned state machine against  test traces and admitting only models whose conformance exceeds a user-defined threshold.

\textbf{Vacuity.} As discussed in Step~6 of \app\ (Section~\ref{sec:step6}), some requirements may be reported as vacuously satisfied. This occurs when a requirement is satisfied because the stimulus that triggers the requirement never occurs. For example, the CTL requirement on line~39 in Figure~\ref{fig:sm2-nusmv} may hold vacuously if the antecedent \texttt{input = attack\_high\_flow \& label = Safe} is unreachable; that is, the IDS is never in the \texttt{Safe} state while high-flow attack traffic occurs.

A vacuity result has two possible interpretations: insufficient environmental coverage or an implicit behavioural property of the SUL. One interpretation is that  vacuity  indicates a limitation of the environment simulator used for data generation. In this case, the simulator may be too restrictive and may fail to generate scenarios in which the IDS is in the \texttt{Safe} state while high-flow attack traffic is present. The vacuity result therefore identifies a gap in the generated data and suggests that the simulator should be revised to exercise the missing scenario. If the simulator is sufficiently expressive, the vacuity result instead reflects an observed characteristic of the SUL behaviour: the IDS never remains in the \texttt{Safe} state under such traffic, because it has already transitioned to another state before the antecedent can hold.

 \section{Evaluation}
\label{sec:eval}
We evaluate \app\ by answering the following research questions (RQs):

\textbf{RQ1 (Complexity and Conformance).} \emph{How effective is the trace abstraction step of \app\ in reducing the complexity of the learned state machines while maintaining a high level of accuracy?} 
In RQ1, we evaluate the impact of \app’s trace abstraction step (Step~3) on the complexity and conformance of the learned state machines. To this end, we compare \app\ with an ablated version in which Step~3 is replaced by an expertise-based numeric abstraction. We then compare \app\ and its ablation in terms of the complexity of the learned models and their conformance to the SUL.

\textbf{RQ2 (Verification).} \emph{Is \app\ effective in verifying requirements for CPS with numeric inputs?} In RQ2, we use \app\ to verify requirements for our case study systems, which are CPS with numeric inputs, and, for each system, present the verification results together with a domain-based validation of the observed outcomes.

\textbf{RQ3 (Exploring Unknown Behaviours).}
\emph{Is \app\ effective in identifying system behaviours that are not explicitly covered by the requirements?} In RQ3, we use the state machines learned by \app\ to explore system behaviours not explicitly specified in the requirements. Specifically, we check temporal-logic properties that capture domain-relevant scenarios, such as how the system behaves under different input conditions or in states not covered by the requirements. We report the observed behaviours together with domain-expert feedback, showing how these findings can help engineers better understand the system.

\subsection{Case Studies}
\label{subsec:CaseStudies}
We use two case-study systems discussed below.

\begin{figure}[t]
    \centering
    \includegraphics[width=\linewidth]{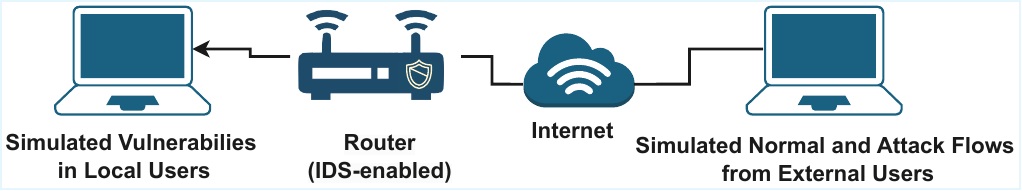}
   \caption{Our testbed simulating the real-world deployment of an IDS-enabled router}
   \label{fig:fig_2}
\end{figure}

\textbf{IDS.} Our first case study is the IDS introduced in Section~\ref{sec:motivation} and detailed in Figure~\ref{fig:conceptual}. To simulate the IDS environment, we use the testbed developed in our previous work in collaboration with RabbitRun Tech. Inc.~\citep{neginconf}. Figure~\ref{fig:fig_2} provides an overview of this testbed. The testbed consists of three virtual machines deployed on separate laptop computers, representing external users, an IDS-enabled router, and local users. It generates normal network traffic together with DoS and DDoS attack traffic, simulating the behaviours of both normal users and attackers among the external users, and uses the \textit{Metasploit} penetration-testing framework~\citep{metasploit} to emulate vulnerabilities in local users.  

\begin{figure}
	\centering
        \includegraphics[width=\linewidth]{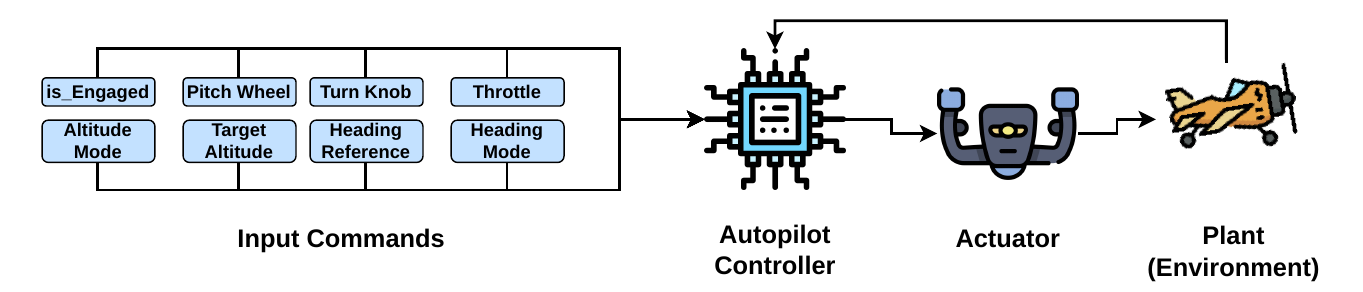}
        \vspace*{-.4cm}
        \caption{Interactions of the autopilot with its environment and the list of input commands provided to the autopilot.}
		\label{fig:ap}
\end{figure}

\textbf{Autopilot system.} Our second system is an autopilot model of a De Havilland Beaver aircraft from a public-domain benchmark of Simulink specifications provided by Lockheed Martin~\citep{lockheedmartin}. Figure~\ref{fig:ap} shows the interactions between the autopilot controller and its environment, including the plant, as well as the list of inputs to the autopilot. When engaged, the autopilot receives commands for throttle, pitch angle, turn rate, heading, and target altitude. It uses these commands to compute control signals, which the actuator converts into physical actions on the aircraft (plant). The resulting aircraft response is fed back to the autopilot, allowing it to continuously adjust its actions toward the target altitude. The Simulink autopilot model includes a simulation environment that, in addition to the plant, models environmental factors such as wind and turbulence.

The autopilot system is expected to satisfy the following requirement: \\$\varphi_3 = $\textit{``When autopilot is engaged, the aircraft should reach a specified altitude within $500$ seconds.''}

Simulations of the autopilot yield three possible outcomes with respect to requirement~$\varphi_3$. The requirement is \emph{unsatisfied} when the aircraft fails to reach the target altitude within the specified time bound. The requirement is \emph{satisfied} when the aircraft reaches the target altitude within the required time bound but maintains it only marginally, with small oscillations around the target altitude under turbulence or other disturbances. The requirement is \emph{robustly satisfied} when the aircraft reaches the target altitude within the required time bound and maintains it consistently, with no significant oscillations around the target altitude despite turbulence or other disturbances.

\begin{figure}
	\centering
        \includegraphics[width=\linewidth]{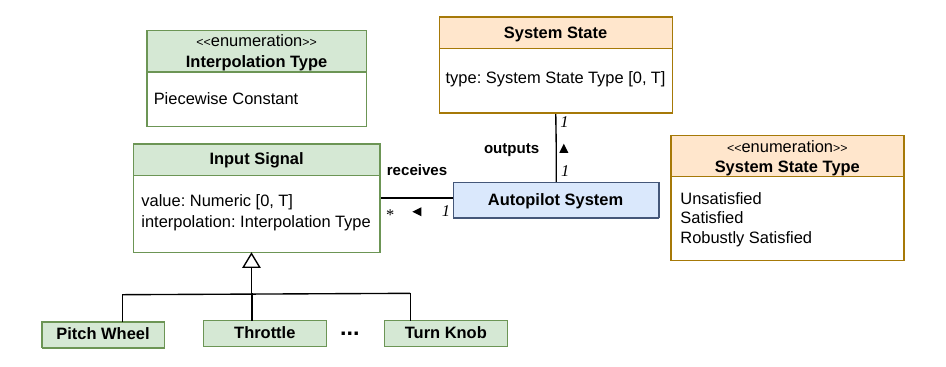}
\caption{A conceptual model for the autopilot system.}
		\label{fig:conceptualap}
\end{figure}

Figure~\ref{fig:conceptualap} shows a conceptual model capturing the inputs and outputs of the autopilot. The inputs are shaded in green, the SUL in blue, and the system-state output in orange.  We generate the autopilot inputs using control-point encoding~\citep{signals19, pareto18}, where each input is defined by values at equally spaced control points. The full time series is then obtained by interpolation. As is common for CPS benchmarks, we use piecewise-constant interpolation~\citep{khandait2024arch, lockheedmartin, cruisecontroller, dcmotor, clutchlockup, guidancecontrol}.

\subsection{RQ1: Complexity and Conformance}
\label{sec:rq1}
Below, we first discuss the experiment design and then present the results obtained for RQ1.

\subsubsection{Ablation}
\label{subsec:baseline}
To answer RQ1, we compare \app\ with an ablation of \app, denoted by \base, that differs from \app\ only in Step~3, trace abstraction. In \base, input selection and range abstraction are guided by domain knowledge rather than learned from the traces. Specifically, the inputs are selected based on domain expertise for the IDS and from system documentation for the autopilot. The numeric range of each selected input is then partitioned into equal-width intervals, while keeping  the number of intervals consistent between \app\ and \base\ for each corresponding system and experiment configuration to ensure a fair comparison. For inconsistency resolution, since \base\ does not learn range abstractions from traces using decision trees, it resolves inconsistencies only by removing inconsistent abstract traces.

\subsubsection{Experiment Design}
\label{sec:expdesign}
We configure \app\ and \base\ using the parameters in Table~\ref{tab:param}. These parameters specify the data generation, trace abstraction, and automata learning settings.

\begin{table}[ht]
\centering
\small
\setlength{\tabcolsep}{3.5pt}
\renewcommand{\arraystretch}{1}
\caption{Parameters for our experiments: (a)~parameters of the learning sets used by \app\ and \base; (b)~parameters of the trace abstraction step of \app; (c)~information about trace abstraction in \base; and (d)~automata learning algorithm used by \app\ and \base.}
\label{tab:param}
\vspace*{-.2cm}
\scriptsize
\begin{tabular}{@{}
p{0.17\columnwidth}
p{0.28\columnwidth}
p{0.10\columnwidth}
p{0.18\columnwidth}
@{}}
\toprule

\multicolumn{4}{c}{\cellcolor{gray!40}\textbf{a. Data Generation for \app\ and \base}} \\ 
\midrule
\textbf{Case-study System} & \textbf{Learning Set} & \textbf{Avg Trace Length} & \textbf{Execution Time (m)} \\ 
\midrule

\multirow{3}{*}{\textbf{IDS}} 
& \texttt{DoS3}: 3-state coverage  & 1530 & 870  \\
& \texttt{DoS5}: 5-state coverage  & 1402 & 474  \\
& \texttt{DDoS}: 3-state coverage  & 1523 & 1420 \\ 
\midrule

\multirow{2}{*}{\textbf{Autopilot}} 
& \texttt{Ascent}: 3-state coverage  & 500 & 125 \\
& \texttt{Descend}: 3-state coverage  & 500 & 125 \\

\midrule
\multicolumn{4}{c}{\cellcolor{gray!40}\textbf{b. Trace Abstraction for \app}} \\ 
\midrule
\textbf{Case-study System} & \textbf{Selected Inputs} 
& \multicolumn{2}{p{0.46\columnwidth}@{}}{\textbf{Range Abstraction by Decision Tree (all case-study systems)}} \\ 
\midrule

\multirow{3}{*}{\textbf{IDS}}
& \texttt{num\_flows} 
& \multicolumn{2}{p{0.46\columnwidth}@{}}{
\multirow{6}{=}{\texttt{Max\_Depth}: 5\newline
\texttt{Purity\_Th}: 0.7\newline
\texttt{Sup\_Th}: $0.2\times n$, where $n$ is the total number of data points
}} \\
& \texttt{num\_unreplied} & \multicolumn{2}{c}{} \\
& \texttt{flow\_unreplied} & \multicolumn{2}{c}{} \\
\cmidrule(lr){1-2}

\multirow{3}{*}{\textbf{Autopilot}}
& \texttt{PitchWheel} & \multicolumn{2}{c}{} \\
& \texttt{Throttle} & \multicolumn{2}{c}{} \\
& \texttt{PWheel\_Throttle} & \multicolumn{2}{c}{} \\

\midrule
\multicolumn{4}{c}{\cellcolor{gray!40}\textbf{c. Trace Abstraction for \base}} \\ 
\midrule
\textbf{Case-study System} & \textbf{Selected Inputs} 
& \multicolumn{2}{p{0.46\columnwidth}@{}}{\textbf{Range Abstraction based on Domain Knowledge (all case-study systems)}} \\ 
\midrule

\multirow{3}{*}{\textbf{IDS}}
& \texttt{num\_flows} 
& \multicolumn{2}{p{0.46\columnwidth}@{}}{
\multirow{6}{=}{Let $d$ be the max range of the numeric variable, and let $k$ be the number of intervals produced by \app. The intervals are
$[0,\Delta)$, $[\Delta,2\times\Delta)$, $\ldots$, $[(k-1)\times\Delta,d]$,
where $\Delta=\frac{d}{k}$.
}} \\
& \texttt{num\_unreplied} & \multicolumn{2}{c}{} \\
& \texttt{flow\_unreplied} & \multicolumn{2}{c}{} \\
\cmidrule(lr){1-2}

\multirow{3}{*}{\textbf{Autopilot}}
& \texttt{PitchWheel} & \multicolumn{2}{c}{} \\
& \texttt{Throttle} & \multicolumn{2}{c}{} \\
& \texttt{PWheel\_Throttle} & \multicolumn{2}{c}{} \\

\midrule
\multicolumn{4}{c}{\cellcolor{gray!40}\textbf{d. Automata Learning for \app\ and \base}} \\ 
\midrule
\textbf{Case-study System} & \multicolumn{3}{c}{\textbf{Passive Learning Algorithm}} \\ 
\midrule

\textbf{IDS} 
& \multicolumn{3}{p{0.74\columnwidth}@{}}{
\multirow{2}{=}{AALpy's generalized state-merging algorithm (GSM)~\cite{muvskardin2022aalpy,gsm}}
} \\ 
\textbf{Autopilot} 
& \multicolumn{3}{c}{} \\

\bottomrule
\end{tabular}

\vspace*{-.1cm}
\end{table} 
\paragraph{Data Generation Setting} We generate time-series data for each case-study system using Step~$1$ in Figure~\ref{fig:mela}, and create traces from this data using Step~$2$. We refer to the resulting trace sets used for automata learning as \emph{learning sets}. In our experiments, both \app\ and \base\ use the same learning sets. For the IDS, we generate time-series vectors over the time domain $[0,256min]$ and use a sampling period of $\delta=10sec$ when converting the time-series vectors into traces. The sampling period matches the minimum refresh rate supported by our IDS testbed. The long time domain was recommended by the domain expert to increase state coverage in the generated traces. We construct separate learning sets for DoS and DDoS attacks so that the learned state machines capture the IDS behaviour under each attack type separately.
To account for randomness in data generation, we repeat Steps~$1$ and~$2$ five times for DoS and five times for DDoS, obtaining five learning sets for each attack type. Among the DoS learning sets, three covered only the states \texttt{Safe}, \texttt{Warning}, and \texttt{Alert}, while the other two covered all five IDS states. For DDoS, all five learning sets covered only \texttt{Safe}, \texttt{Warning}, and \texttt{Alert}.  As we will discuss in Section~\ref{sec:rq3}, the domain expert confirmed that \texttt{Tending Warning} and \texttt{Tending Alert} are unreachable under DDoS attacks.

For the autopilot system, we generate time-series vectors over the time domain $[0,500sec]$ and sample them with $\delta=1sec$ during trace creation. The $500$-second duration matches requirement $\varphi_3$ in Section~\ref{subsec:CaseStudies}, which specifies a 500-second time bound, and the sampling period matches the time resolution used for data generation in the Simulink model.
We construct separate learning sets for the two cases of requirement $\varphi_3$: ascent to a target altitude and descent to a target altitude. To account for randomness in data generation, we repeat Steps~$1$ and~$2$ fifteen times for each case, obtaining fifteen learning sets for ascent and fifteen learning sets for descent.

In summary, we use the following learning sets for our experiments. For the IDS, we construct three learning sets by taking the union of learning sets with the same attack type and state coverage: \texttt{DoS3} is the union of the three learning sets for DoS with three-state coverage, \texttt{DoS5} is the union of the two learning sets for DoS with five-state coverage, and \texttt{DDoS} is the union of the five learning sets for DDoS. For the autopilot, \texttt{Ascent} combines the fifteen learning sets generated for ascent to a target altitude, and \texttt{Descent} combines the fifteen learning sets generated for descent to a target altitude. Table~\ref{tab:param}(a) shows the average length of traces for each learning set as well as the total execution time (in minutes) required to generate the traces in these learning sets.

\paragraph{Trace Abstraction Setting}
We configure the trace-abstraction step, i.e., Step~$3$ in Figure~\ref{fig:mela}, separately for \app\ and \base. In both approaches, trace abstraction consists of input selection, range abstraction, and inconsistency resolution.

\textbf{Settings for \app.} Table~\ref{tab:param}(b) summarizes the trace abstraction setting for \app. For \emph{input selection}, the requirement-referenced input for the IDS is \texttt{External User type}, which distinguishes normal traffic from attack traffic. In addition to this input, \app\ considers \texttt{num\_flows} and \texttt{num\_unreplied}, the two inputs whose information-gain values exceed the standard deviation computed over the information-gain values of all inputs, as discussed in Section~\ref{sec:varsel}. For each IDS learning set, we define three configurations, each including \texttt{External User type} and one of the following choices of additional inputs: \texttt{num\_flows}, \texttt{num\_unreplied}, or both. We refer to these configurations as \texttt{num\_flows}, \texttt{num\_unreplied}, and \texttt{flow\_unreplied}, respectively.

For the autopilot, the requirement-referenced input is \texttt{Target Altitude}. Among the remaining inputs, \app\ identifies \texttt{PitchWheel} and \texttt{Throttle} as the inputs with the highest information-gain values. We define three autopilot configurations, each including \texttt{Target Altitude}: \texttt{PitchWheel}, \texttt{Throttle}, and \texttt{PWheel\_Throttle}. These configurations use \texttt{PitchWheel}, \texttt{Throttle}, and both inputs, respectively.

For \emph{range abstraction}, we configure the decision trees used to abstract the numeric inputs of each case-study system according to the parameters in Table~\ref{tab:param}(b). We set the maximum tree depth to five, denoted by \texttt{Max\_Depth}, to limit overfitting and avoid generating many leaves that would split numeric domains into overly fine-grained intervals. We set the support threshold to $20\%$ of the data used to build the tree, denoted by \texttt{Sup\_Th}, and the purity threshold to $70\%$, denoted by \texttt{Purity\_Th}. These thresholds ensure that the leaves selected for interval construction are both sufficiently populated and strongly associated with a specific system state.

For \emph{inconsistency resolution}, \app\ first refines the abstraction and removes traces only when refinement cannot eliminate the remaining inconsistencies. In our case studies, range-abstraction refinement resolved most inconsistencies, and only $0.1\%$ of traces had to be removed.

\textbf{Settings for \base.} Table~\ref{tab:param}(c) summarizes the trace-abstraction settings for \base. 
For \emph{input selection}, \base\ selects inputs based on domain knowledge. For the IDS, the domain expert identified \texttt{num\_flows} and \texttt{num\_unreplied} as the inputs that most strongly affect the IDS state. For the autopilot, we used the system documentation~\citep{federal2009pilot, autopilothandbook} and empirical results from the literature~\citep{TOSEM, TSE} to identify \texttt{PitchWheel} and \texttt{Throttle} as the inputs that affect altitude and, consequently, the system state. These inputs match those selected by \app. Therefore, for \base, we use the same input-selection configurations defined for \app.

For \emph{range abstraction}, each selected numeric input is partitioned into the same number of equal-width intervals as in \app\ for the corresponding system and learning-set configuration. This keeps the abstraction granularity identical between \app\ and \base. Furthermore, using the same number of equal-width intervals in \base\ controls for abstraction granularity, ensuring that any differences from \app\ are due to how interval boundaries are derived rather than to the number of abstract input symbols. For \emph{inconsistency resolution}, as mentioned in Section~\ref{subsec:baseline}, \base\ resolves inconsistencies only by removing inconsistent traces. In our experiments, only $0.1\%$ of \base\ traces had to be removed.

\paragraph{Automata Learning} As shown in Table~\ref{tab:param}(d), both \app\ and \base\ use AALpy's implementation of generalized state merging (GSM), an RPNI-based passive learning algorithm~\citep{muvskardin2022aalpy,gsm} to produce state machines.

\subsubsection{Metrics}
\label{subsec:metric}
To assess the complexity of the generated state machines, we report the number of states, the number of transitions, and the size of the input alphabet. These three size-based metrics are commonly used in the literature to evaluate the complexity of state machines~\citep{Hall11}.

We measure the accuracy (or conformance) of the generated state machines using the conformance\hyp{}checking process in Step~5 of \app, which evaluates whether each learned automaton accepts unseen test traces  from the SUL. For the IDS, we generate five test sets to cover a range of trace lengths: \emph{very small}, \emph{small}, \emph{medium}, \emph{large}, and \emph{very large}. Each set contains $100$ traces, resulting in $500$ test traces in total. The average trace lengths are $70$, $138$, $207$, $276$, and $345$, corresponding to approximately $10\%$, $20\%$, $30\%$, $40\%$, and $50\%$ of the learning-trace length, respectively. Because generating long IDS traces is costly, we cap the test-trace length at $50\%$ of the learning-trace length.

For the autopilot, each test trace has length $500$, similar to the traces in the learning set, since requirement $\varphi_3$ in Section~\ref{subsec:CaseStudies} requires simulating the system for $500$ seconds to observe the altitude change. We therefore generate two test sets, one for ascent to the target altitude and one for descent to the target altitude, with six traces in each set.

\subsubsection{Results}
To answer RQ1, we apply \app\ and \base\ to the IDS learning sets (\texttt{DoS3}, \texttt{DoS5}, and \texttt{DDoS}) and the autopilot learning sets (\texttt{Ascent} and \texttt{Descent}).  Across these learning sets, we generate 30 state machines by applying the two approaches under three abstraction configurations for each case study: \texttt{num\_flows}, \texttt{num\_unreplied}, and \texttt{flow\_unreplied} for the IDS, and \texttt{PitchWheel}, \texttt{Throttle}, and \texttt{PWheel\_Throttle} for the autopilot.

\begin{table}[t]
\centering
\caption{Comparing the number of states, number of transitions, and the alphabet size for the state machines learned by \app\ versus by \base.}
\label{tab:tbl_1}

\scriptsize
\setlength{\tabcolsep}{3pt}
\renewcommand{\arraystretch}{1.08}

\begin{tabular}{|C{0.14\columnwidth}|C{0.105\columnwidth}|L{0.195\columnwidth}|C{0.062\columnwidth}|C{0.075\columnwidth}|C{0.068\columnwidth}|C{0.062\columnwidth}|C{0.075\columnwidth}|C{0.068\columnwidth}|}
\hhline{|---------|}
\multirow{2}{*}{\makecell{\textbf{Case}\\\textbf{study}}} &
\multirow{2}{*}{\makecell{\textbf{Learning}\\\textbf{set}}} &
\multirow{2}{*}{\centering\arraybackslash\textbf{Configuration}} &
\multicolumn{3}{c|}{\textbf{\app}} &
\multicolumn{3}{c|}{\textbf{\base}}  \\
\hhline{|~|~|~|------|}
 & & &
\rotatebox{90}{\textbf{\# States}} &
\rotatebox{90}{\textbf{\# Transitions}} &
\rotatebox{90}{$\mathbf{\lvert Alphabet\rvert}$} &
\rotatebox{90}{\textbf{\# States}} &
\rotatebox{90}{\textbf{\# Transitions}} &
\rotatebox{90}{$\mathbf{\lvert Alphabet\rvert}$} \\
\hhline{|---------|}

\multirow{9}{*}{\textbf{IDS}} &
\multirow{3}{*}{\textbf{\texttt{DoS3}}} & \texttt{num\_flows} & 3 & 9 & 5 & 290 & 325 & 5\\
\hhline{|~|~|-------|}
 &  & \texttt{num\_unreplied} & 19 & 51 & 5 & 29 & 68 & 5\\
\hhline{|~|~|-------|}
 &  & \texttt{flow\_unreplied} & 3 & 11 & 7 & 23 & 65 & 8 \\
\hhline{|~|--------|}

&
\multirow{3}{*}{\textbf{\texttt{DDoS}}} & \texttt{num\_flows} & 3 & 9 & 5 & 284 & 319 & 5 \\
\hhline{|~|~|-------|}
 &  & \texttt{num\_unreplied} & 17 & 46 & 5 & 24 & 62 & 5 \\
\hhline{|~|~|-------|}
 &  & \texttt{flow\_unreplied} & 3 & 11 & 7 & 20 & 52 & 8 \\
\hhline{|~|--------|}

&
\multirow{3}{*}{\textbf{\texttt{DoS5}}} & \texttt{num\_flows} & 47 & 78 & 5 & 410 & 499 & 5 \\
\hhline{|~|~|-------|}
 &  & \texttt{num\_unreplied} & 121 & 187 & 5 & 178 & 267 & 5 \\
\hhline{|~|~|-------|}
 &  & \texttt{flow\_unreplied} & 46 & 83 & 8 & 167 & 253 & 9 \\
\hhline{|=========|"}

\multirow{6}{*}{\textbf{Autopilot}} &
\multirow{3}{*}{\textbf{\texttt{Descend}}} & \texttt{PitchWheel} & 40 & 65 & 3 & 59 & 86 & 2 \\
\hhline{|~|~|-------|}
 &  & \texttt{Throttle} & 126 & 169 & 2 & 178 & 178 & 1 \\
\hhline{|~|~|-------|}
 &  & \texttt{PWheel\_Throttle} & 40 & 75 & 6 & 59 & 86 & 2 \\
\hhline{|~|--------|}

&
\multirow{3}{*}{\textbf{\texttt{Ascent}}} & \texttt{PitchWheel} & 53 & 106 & 3 & 76 & 135 & 3 \\
\hhline{|~|~|-------|}
 &  & \texttt{Throttle} & 133 & 163 & 2 & 180 & 180 & 1 \\
\hhline{|~|~|-------|}
 &  & \texttt{PWheel\_Throttle} & 53 & 125 & 4 & 76 & 135 & 3 \\
\hhline{|---------|}

\end{tabular}
\end{table} 
To assess complexity, Table~\ref{tab:tbl_1} reports the number of states, transitions, and input-alphabet sizes for the state machines generated by \app\ and \base\ for each case study. Overall, the state machines generated by \app\ are consistently more compact than those generated by \base, with fewer states and transitions across both the IDS and autopilot case studies. 
Across all configurations, the ML-based trace abstraction used in \app\ reduces the number of states by an average of $53.79$\% and the number of transitions by an average of $44.62$\% compared to the expertise-based abstraction used in \base.

Regarding input-alphabet size, the state machines generated by \app\ for the IDS use alphabets that are either the same size as or smaller than those of the corresponding \base\ state machines. For the autopilot, one of the six state machines generated by \app\ has the same alphabet size as its \base\ counterpart, while the remaining five use larger alphabets due to the finer-grained abstractions derived by \app.

\begin{figure}[!ht]
    \centering

    \begin{subfigure}[t]{0.70\linewidth}
        \centering
        \includegraphics[width=\linewidth]{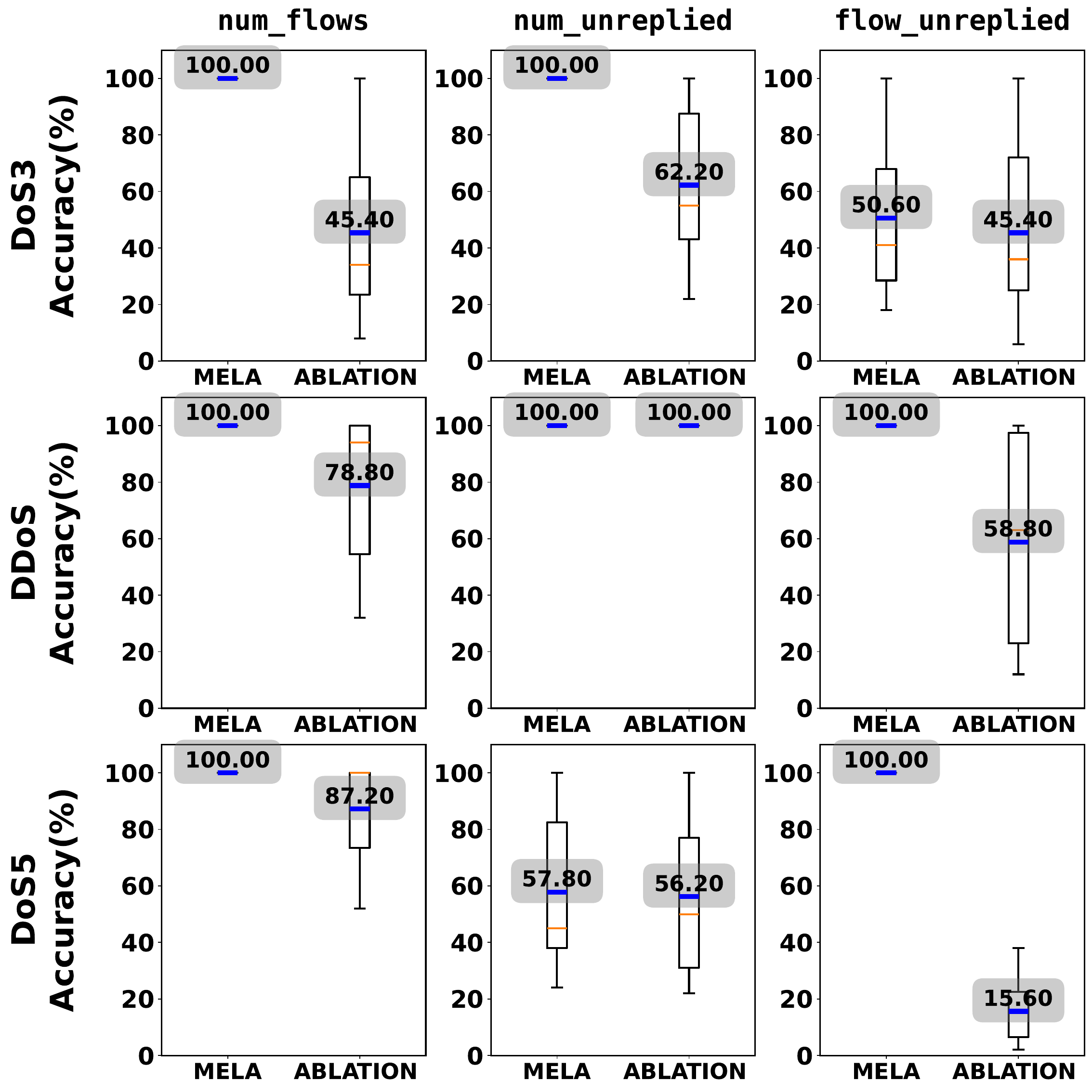}
        \caption{IDS case study.}
        \label{fig:boxplot_IDS}
    \end{subfigure}
    \hfill
    \begin{subfigure}[t]{0.70\linewidth}
        \centering
        \includegraphics[width=\linewidth]{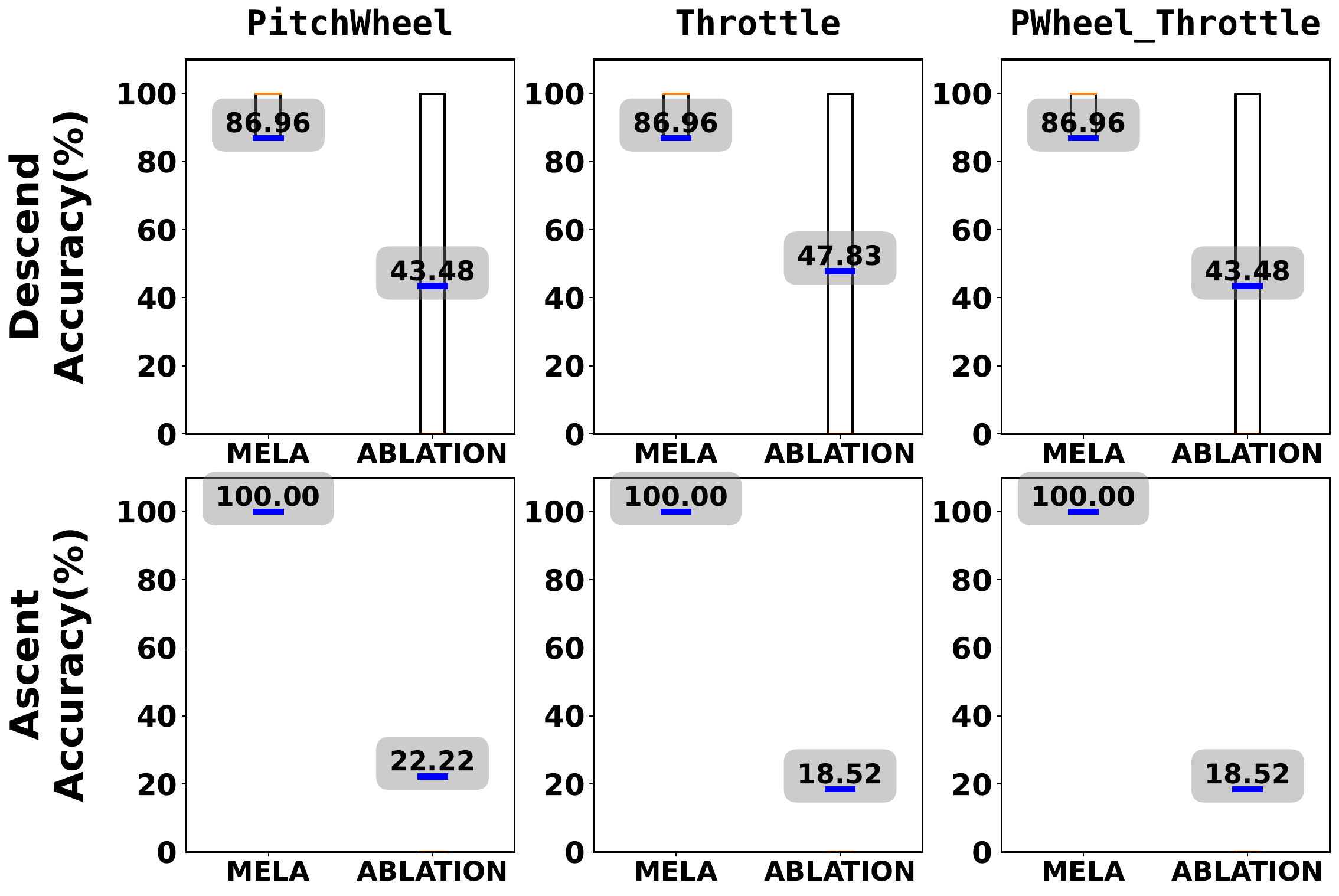}
        \caption{Autopilot case study.}
        \label{fig:boxplot_AT}
    \end{subfigure}

    \vspace*{-.2cm}
    \caption{Comparing the accuracy of the state machines learned by \app\ versus by \base. 
    (a) shows the results for the IDS learning sets (\texttt{DoS3}, \texttt{DoS5}, and \texttt{DDoS}). 
    (b) shows the results for the Autopilot learning sets (\texttt{Ascend} and \texttt{Descend}). 
    Both case studies are evaluated under different configurations for selected inputs defined in Table~\ref{tab:param}.}
    \label{fig:boxplot}
    \vspace*{-.3cm}
\end{figure}

Figure~\ref{fig:boxplot} shows the accuracy results for the learned automata by \app\ and \base.  The accuracy values are computed by applying each of the $30$ learned automata to the test traces described in Section~\ref{subsec:metric}. The plots in the figure show the accuracy distribution of each learned automaton with respect to these test traces. Figure~\ref{fig:boxplot}(a) shows the accuracy results for the IDS. Of the nine state machines generated by \app, eight have higher average accuracy than the corresponding state machines generated by \base, and one has the same average accuracy. Seven of the nine IDS state machines learned by \app\ reach $100\%$ accuracy. Figure~\ref{fig:boxplot}(b) shows the accuracy results for the autopilot. All six state machines learned by \app\ are more accurate than their corresponding \base\ state machines. The three state machines learned by \app\ from the \texttt{Ascent} learning set reach $100\%$ accuracy, while the three learned from the \texttt{Descent} learning set reach $86.96\%$ accuracy. Overall, across both case studies, the state machines learned by \app\ are, on average, $41.71\%$ more accurate than those learned by \base.

To assess whether these differences are statistically significant, we use the Mann-Whitney U test with a significance level of $0.05$~\citep{Mann-Whitney} and Vargha-Delaney's effect size $\hat{A}_{12}$~\citep{vargha2000critique}. We compute $\hat{A}_{12}$ from the accuracy values, where $\hat{A}_{12}>0.5$ indicates that \app\ achieves higher accuracy than \base, while $\hat{A}_{12}<0.5$ indicates the opposite. We classify effect size values as small, medium, and large when their values are greater than or equal to $0.56$, $0.64$, and $0.71$, respectively~\citep{vargha2000critique}; values between $0.44$ and $0.56$ are considered negligible. Statistical tests comparing the accuracy results in Figures~\ref{fig:boxplot}(a) and~(b) are provided in Table~\ref{tab:stat-test}. Each test compares the accuracy of an automaton generated by \app\ with the corresponding automaton generated by \base\ for the same learning set and variable-selection configuration. As shown in the table, \app\ achieves significantly higher accuracy than \base\ in nine of the 15 comparisons. The effect sizes are large for the autopilot system and medium or large for the IDS. For the remaining six comparisons, the accuracy differences between \app\ and \base\ are not statistically significant.

\begin{table}[t]
\centering
\captionsetup[table]{labelfont=bf}
\caption{Statistical tests comparing the accuracy results of \app\ against those of \base. The p-values highlighted in blue represent cases where \app\ significantly outperforms \base. The significance level is $0.05$.
N, S, M, and L indicate a negligible, small, medium and large effect sizes, respectively.}
\label{tab:stat-test}

\begin{tabular}{|c|c|c|c|c|}
\hhline{|-----|}
\textbf{Case study} & \textbf{Learning set} & \textbf{Configuration} & \textbf{$p$-value} & \textbf{$\hat{A}_{12}$} \\
\hhline{|-----|}

\multirow{9}{*}{\textbf{IDS}}
 & \multirow{3}{*}{\textbf{\texttt{DoS3}}}
 & \texttt{num\_flows}   & \cellcolor[HTML]{96FFFB}0.000 & 0.95 (L) \\ 
\hhline{|~|~|---|}
 &       & \texttt{num\_unreplied}   & \cellcolor[HTML]{96FFFB}0.001 & 0.90 (L) \\ 
\hhline{|~|~|---|}
 &       & \texttt{flow\_unreplied} & 0.762 & 0.55 (N) \\
\hhline{|~|----|}

 & \multirow{3}{*}{\textbf{\texttt{DoS5}}}
 & \texttt{num\_flows}   & \cellcolor[HTML]{96FFFB}0.035 & 0.70 (M) \\ 
\hhline{|~|~|---|}
 &       & \texttt{num\_unreplied}   & 0.676 & 0.44 (N) \\ 
\hhline{|~|~|---|}
 &       & \texttt{flow\_unreplied} & \cellcolor[HTML]{96FFFB}0.001 & 1.00 (L) \\
\hhline{|~|----|}

 & \multirow{3}{*}{\textbf{\texttt{DDoS}}}
 & \texttt{num\_flows}   & \cellcolor[HTML]{96FFFB}0.015 & 0.75 (L) \\ 
\hhline{|~|~|---|}
 &       & \texttt{num\_unreplied}   & 1.000 & 0.50 (N) \\ 
\hhline{|~|~|---|}
 &       & \texttt{flow\_unreplied} & \cellcolor[HTML]{96FFFB}0.002 & 0.85 (L) \\
\hhline{|=====|}

\multirow{6}{*}{\textbf{Autopilot}}
 & \multirow{3}{*}{\textbf{\texttt{Descend}}}
 & \texttt{PitchWheel}   & 0.050 & 0.82 (L) \\ 
\hhline{|~|~|---|}
 &       & \texttt{Throttle}   & 0.181 & 0.72 (L) \\ 
\hhline{|~|~|---|}
 &       & \texttt{PWheel\_Throttle} & 0.050 & 0.82 (L) \\
\hhline{|~|----|}

 & \multirow{3}{*}{\textbf{\texttt{Ascent}}}
 & \texttt{PitchWheel}   & \cellcolor[HTML]{96FFFB} 0.025 & 0.83 (L) \\ 
\hhline{|~|~|---|}
 &       & \texttt{Throttle}   & \cellcolor[HTML]{96FFFB} 0.007 & 0.92 (L) \\ 
\hhline{|~|~|---|}
 &       & \texttt{PWheel\_Throttle} & \cellcolor[HTML]{96FFFB} 0.007 & 0.92 (L) \\
\hhline{|-----|}
\end{tabular}
\end{table}
 
\begin{tcolorbox}[breakable,colback=gray!10!white,colframe=black!75!black]
\textbf{Finding.} Our approach (\app) leads to an average reduction of $49.20\%$ in the number of states and transitions of the learned automata, while improving accuracy by an average of $41.71\%$ compared to using expertise-based abstractions for automata learning. Statistical-test results show that this accuracy improvement is significant in nine of the $15$ comparisons, with medium or large effect sizes.
\end{tcolorbox}

\subsection{RQ2: Verification}
\label{sec:rq2}
We discuss the experiment design and the results obtained for RQ2.

\subsubsection{Experiment Design}
For this research question, we reuse the state machines generated in RQ1. For the IDS, we analyze the models learned from the \texttt{DDoS} learning set and from \texttt{DoS5}. We choose \texttt{DoS5} instead of \texttt{DoS3} because \texttt{DoS5} exercises all five IDS states under DoS attacks, while \texttt{DoS3} does not. For \texttt{DoS5}, we keep the models built with \texttt{num\_flows} and \texttt{flow\_unreplied}, as both show high conformance accuracy, and discard the \texttt{num\_unreplied} model because its accuracy is relatively lower (Figure~\ref{fig:boxplot}). To simplify the presentation of RQ2 and RQ3, we refer to the three IDS abstraction configurations based on \texttt{num\_flows}, \texttt{flow\_unreplied}, and \texttt{num\_unreplied} as \texttt{config1}, \texttt{config2}, and \texttt{config3}, respectively. For the autopilot, we retain all RQ1 state machines because all configurations achieve high accuracy, above $85\%$ (Figure~\ref{fig:boxplot}).

We decompose each IDS requirement into three properties because $\varphi_1$ and $\varphi_2$ specify staged behaviour. For $\varphi_1$, we separately check the progression from \texttt{Safe} to \texttt{Warning}, from \texttt{Warning} to \texttt{Alert}, and persistence in \texttt{Alert} under continued attacks; $\varphi_2$ is decomposed analogously for staged recovery after attacks stop.
Similarly, we decompose the autopilot requirement, i.e., $\varphi_3$ from Section~\ref{subsec:CaseStudies}, into two properties to separately capture the two transitions in satisfaction status: from unsatisfied to satisfied, and from satisfied to robustly satisfied.
We encode the detailed  requirements in CTL. For readability, the leftmost columns of Tables~\ref{tab:main-table}(a) and~(b) and Table~\ref{tab:ap-self-loop-results} provide their natural-language requirements for the IDS and autopilot case studies, respectively. 
For example, the first row of Table~\ref{tab:main-table}(a) corresponds to the CTL property on line~$39$ of Figure~\ref{fig:sm2-nusmv}. The full set of CTL properties, including the autopilot descent case, is available in our supplementary material~\citep{MELARepoEvaluation,MELARepoResult}. In Table~\ref{tab:main-table}, we abbreviate the IDS states \texttt{Safe}, \texttt{Warning}, \texttt{Alert}, \texttt{Tending Warning}, and \texttt{Tending Alert} as ``S'', ``W'', ``A'', ``TW'', and ``TA'', respectively. In Tables~\ref{tab:main-table} and~\ref{tab:ap-self-loop-results}, the symbols \checkmark, $\times$, and v denote that \app\ reports a property as satisfied, violated, or vacuously satisfied, respectively.

\begin{table}[!ht]
\centering
\caption{Verification results for properties derived from the IDS requirements $\varphi_1$ and $\varphi_2$. The properties related to $\varphi_1$ are evaluated under different attack-traffic intensities, while the properties related to $\varphi_2$ are evaluated under different normal-traffic intensities. The table shows whether each property is satisfied (\checkmark), violated ($\times$), or vacuously satisfied (v).}
\label{tab:main-table}
\footnotesize
\setlength{\tabcolsep}{4pt}
\renewcommand{\arraystretch}{1.08}

\begin{tabular}{|p{0.48\linewidth}|C{0.06\linewidth}C{0.06\linewidth}C{0.06\linewidth}|C{0.06\linewidth}C{0.06\linewidth}C{0.06\linewidth}|}
\multicolumn{7}{@{}p{\linewidth}@{}}{\textbf{(a) Results for state machines learned from the \texttt{DoS5} learning set, capturing IDS behaviour under DoS attacks.}} \\
\hline
\multicolumn{1}{|c|}{\multirow{2}{*}{\textbf{Property}}} &
\multicolumn{3}{c|}{\textbf{\texttt{config1}}} &
\multicolumn{3}{c|}{\textbf{\texttt{config3}}} \\
\hhline{|~|---|---|}
& \multicolumn{1}{>{\columncolor[HTML]{D9EAD3}}C{0.06\linewidth}}{\textbf{Low}}
& \multicolumn{1}{>{\columncolor[HTML]{FCE5CD}}C{0.06\linewidth}}{\textbf{Med}}
& \multicolumn{1}{>{\columncolor[HTML]{EADCF8}}C{0.06\linewidth}|}{\textbf{High}}
& \multicolumn{1}{>{\columncolor[HTML]{D9EAD3}}C{0.06\linewidth}}{\textbf{Low}}
& \multicolumn{1}{>{\columncolor[HTML]{FCE5CD}}C{0.06\linewidth}}{\textbf{Med}}
& \multicolumn{1}{>{\columncolor[HTML]{EADCF8}}C{0.068\linewidth}|}{\textbf{High}} \\
\hline

\multicolumn{1}{|c|}{\cellcolor{gray!15}\textbf{Properties related to $\varphi_1$}} &
\multicolumn{6}{c|}{\cellcolor{gray!15}\textbf{Attack traffic intensity (Low / Med / High)}} \\
\hline

When attacks happen and the system is in state \texttt{S}, the system shall transition to \texttt{W} or \texttt{TW}.
& $\times$ & \checkmark & v
& $\times$ & \checkmark & v \\
\hline

When attacks happen and the system is in state \texttt{W}, the system shall transition to \texttt{A} or \texttt{TA}.
& v & $\times$ & \checkmark
& v & $\times$ & \checkmark \\
\hline

When attacks happen and the system is in state \texttt{A}, the system shall remain in \texttt{A} or transition to \texttt{TA}.
& v & $\times$ & \checkmark
& v & $\times$ & \checkmark \\
\hline

\multicolumn{1}{|c|}{\cellcolor{gray!15}\textbf{Properties related to $\varphi_2$}} &
\multicolumn{6}{c|}{\cellcolor{gray!15}\textbf{Normal traffic intensity (Low / Med / High)}} \\
\hline

When attacks are stopped and the system is in state \texttt{S}, the system shall remain in \texttt{S}.
& \checkmark & v & v
& \checkmark & v & v \\
\hline

When attacks are stopped and the system is in state \texttt{W}, the system shall transition to \texttt{S}.
& \checkmark & $\times$ & v
& \checkmark & $\times$ & v \\
\hline

When attacks are stopped and the system is in state \texttt{A}, the system shall transition to \texttt{W} or \texttt{TW}.
& v & v & v
& v & v & v \\
\hline
\end{tabular}

\vspace*{.4cm}

\setlength{\tabcolsep}{2pt}

\begin{tabular}{|p{0.46\linewidth}|C{0.045\linewidth}C{0.045\linewidth}C{0.05\linewidth}|C{0.045\linewidth}C{0.045\linewidth}C{0.05\linewidth}|C{0.045\linewidth}C{0.045\linewidth}C{0.05\linewidth}|}
\multicolumn{10}{@{}p{\linewidth}@{}}{\textbf{(b) Results for state machines learned from the \texttt{DDoS} learning set, capturing IDS behaviour under DDoS attacks.}} \\
\hline
\multicolumn{1}{|c|}{\multirow{2}{*}{\textbf{Property}}} &
\multicolumn{3}{c|}{\textbf{\texttt{config1}}} &
\multicolumn{3}{c|}{\textbf{\texttt{config2}}} &
\multicolumn{3}{c|}{\textbf{\texttt{config3}}} \\
\hhline{|~|---|---|---|}
& \multicolumn{1}{>{\columncolor[HTML]{D9EAD3}}C{0.045\linewidth}}{\textbf{Low}}
& \multicolumn{1}{>{\columncolor[HTML]{FCE5CD}}C{0.045\linewidth}}{\textbf{Med}}
& \multicolumn{1}{>{\columncolor[HTML]{EADCF8}}C{0.05\linewidth}|}{\textbf{High}}

& \multicolumn{1}{>{\columncolor[HTML]{D9EAD3}}C{0.045\linewidth}}{\textbf{Low}}
& \multicolumn{1}{>{\columncolor[HTML]{FCE5CD}}C{0.045\linewidth}}{\textbf{Med}}
& \multicolumn{1}{>{\columncolor[HTML]{EADCF8}}C{0.05\linewidth}|}{\textbf{High}}

& \multicolumn{1}{>{\columncolor[HTML]{D9EAD3}}C{0.045\linewidth}}{\textbf{Low}}
& \multicolumn{1}{>{\columncolor[HTML]{FCE5CD}}C{0.045\linewidth}}{\textbf{Med}}
& \multicolumn{1}{>{\columncolor[HTML]{EADCF8}}C{0.055\linewidth}|}{\textbf{High}} \\
\hline

\multicolumn{1}{|c|}{\cellcolor{gray!15}\textbf{Properties related to $\varphi_1$}} &
\multicolumn{9}{c|}{\cellcolor{gray!15}\textbf{Attack traffic intensity (Low / Med / High)}} \\
\hline

When attacks happen and the system is in state \texttt{S}, the system shall transition to \texttt{W} or \texttt{TW}.
& $\times$ & \checkmark & v
& $\times$ & \checkmark & v
& $\times$ & \checkmark & v \\
\hline

When attacks happen and the system is in state \texttt{W}, the system shall transition to \texttt{A} or \texttt{TA}.
& v & $\times$ & \checkmark\textsuperscript{1}
& v & $\times$ & $\times$\textsuperscript{1}
& v & $\times$ & \checkmark\textsuperscript{1} \\
\hline

When attacks happen and the system is in state \texttt{A}, the system shall remain in \texttt{A} or transition to \texttt{TA}.
& v & $\times$ & \checkmark
& v & $\times$ & \checkmark
& v & $\times$ & \checkmark \\
\hline

\multicolumn{1}{|c|}{\cellcolor{gray!15}\textbf{Properties related to $\varphi_2$}} &
\multicolumn{9}{c|}{\cellcolor{gray!15}\textbf{Normal traffic intensity (Low / Med / High)}} \\
\hline

When attacks are stopped and the system is in state \texttt{S}, the system shall remain in \texttt{S}.
& \checkmark & v & v
& \checkmark & v & v
& \checkmark & v & v \\
\hline

When attacks are stopped and the system is in state \texttt{W}, the system shall transition to \texttt{S}.
& \checkmark\textsuperscript{2} & $\times$ & v
& $\times$\textsuperscript{2} & $\times$ & v
& \checkmark\textsuperscript{2} & $\times$ & v \\
\hline

When attacks are stopped and the system is in state \texttt{A}, the system shall transition to \texttt{W} or \texttt{TW}.
& v & v & v
& v & v & v
& v & v & v \\
\hline

\multicolumn{10}{p{0.96\linewidth}}{\footnotesize
\textsuperscript{1,2} Non-agreement combinations: for the same property, learning set, and input range, the three configurations do not produce the same verification outcome. In both marked combinations, \texttt{config1} and \texttt{config3} produce \checkmark, while \texttt{config2} produces $\times$.} \\

\end{tabular}

\end{table}

\subsubsection{Results}\mbox{}\\
\textbf{IDS.} Tables~\ref{tab:main-table}(a) and~(b) show the verification results for the IDS using the \texttt{DoS5} and \texttt{DDoS} learning sets, respectively. Based on Tables~\ref{tab:main-table}(a) and~(b), the IDS reacts to DoS/DDoS attacks only when the number of flows is medium or high, while low-flow attacks do not trigger a reaction. Medium-flow attacks prompt the IDS to transition from S to W or TW, but do not move it further to the critical states, TA or A. Only high-flow attacks move the IDS from W to A or TA, and once the IDS reaches A, it stays in A or TA.
Further, the results show that the IDS returns to state S only with low-flow normal traffic. In addition, the combination of normal traffic with a high number of flows does not appear in state machines learned by \app. The IDS remains in the W state for medium-flow normal traffic, indicating that it considers this traffic suspicious.

In addition, the verification results in Tables~\ref{tab:main-table}(a) 
and~(b) are largely consistent across the three configurations,  \texttt{config1}, \texttt{config2}, and \texttt{config3}. We define a  combination as a specific property, learning set, and input range. A combination 
is consistent when all three configurations produce the same verification outcome.  Across the $36$ combinations in Tables~\ref{tab:main-table}(a) and~(b), $34$ are 
consistent. The two remaining combinations, marked by superscripts $1$ and $2$  in Table~\ref{tab:main-table}(b), are non-agreement combinations. In both,  \texttt{config1} and \texttt{config3} produce \checkmark, whereas  \texttt{config2} produces $\times$. Thus, even in the two non-agreement  combinations, the verification outcome is supported by two of the three learned 
state machines.

\textbf{Domain-based validation for IDS.} The results in Tables~\ref{tab:main-table}(a) and~(b) are consistent with the domain expert's assessment of the IDS behaviour. The expert confirmed that low-flow DoS and DDoS attacks do not degrade the quality of service for clients; therefore, the IDS is not expected to escalate in response to them. The absence of high-flow normal traffic from the learned state machines matches the expert's intuition that high numbers of flows are generated by attackers rather than normal users. Finally, the expert confirmed that DoS and DDoS attacks typically begin with a low number of flows and gradually escalate, which explains why high-flow attacks are not observed while the IDS is in state \texttt{S}.

\textbf{Autopilot.} Table~\ref{tab:ap-self-loop-results} summarizes the verification results for the autopilot in the \texttt{Ascent} scenario. The results indicate that requirement~$\varphi_3$ is not satisfied when \texttt{PitchWheel} remains low over time. A low or high \texttt{Throttle} can lead to the satisfaction of $\varphi_3$, but robust satisfaction occurs only when \texttt{Throttle} is high. When the state machine includes both \texttt{PitchWheel} and \texttt{Throttle}, \texttt{PitchWheel} plays the dominant role: $\varphi_3$ is satisfied or robustly satisfied when \texttt{PitchWheel} is medium and low \texttt{Throttle} is applied. When both inputs are high, robust satisfaction is ensured. Applying a high \texttt{PitchWheel} with any \texttt{Throttle} value ensures satisfaction.

For the \texttt{Descent} learning set, whose verification results are provided in our online repository~\citep{descentrq2}, the results show, as expected, the reverse pattern for \texttt{PitchWheel}: since a low \texttt{PitchWheel} command drives the aircraft toward a lower altitude, requirement~$\varphi_3$ is satisfied or robustly satisfied only when \texttt{PitchWheel} remains low over time. This pattern also holds when the state machine includes both \texttt{PitchWheel} and \texttt{Throttle}: satisfaction of $\varphi_3$ in the descent scenario depends on \texttt{PitchWheel} remaining low, while the \texttt{Throttle} level does not affect the outcome.

\textbf{Domain-based validation for autopilot.} The verification results for both \texttt{Ascent} and \texttt{Descent} learning sets are consistent with the system documentation: To satisfy requirement $\varphi_3$, the aircraft must maintain the proper nose attitude and receive sufficient throttle input~\citep{federal2009pilot, autopilothandbook}. For ascent, satisfying $\varphi_3$ requires the aircraft to maintain an upward nose orientation while receiving sufficient engine power to sustain the climb. Through trace abstraction, \app\ identifies upward-nose behaviour with medium or high \texttt{PitchWheel} values, whereas \texttt{Throttle} mainly determines whether satisfaction is robust.
The vacuous verification results for the ascent scenario indicate that high \texttt{Throttle} can override the effect of low or medium \texttt{PitchWheel} values. Specifically, high throttle can drive the aircraft nose into a strongly upward orientation, i.e., high \texttt{PitchWheel}, even when the initially supplied \texttt{PitchWheel} value is only slightly or moderately upward.
For descent, the documentation indicates that the aircraft nose must remain oriented downward to reduce altitude, which corresponds to low \texttt{PitchWheel} in the learned abstraction.

\begin{tcolorbox}[breakable,colback=gray!10!white,colframe=black!75!black]
\textbf{Finding.}
Overall, our results show that the state machines learned by \app\ are effective for verifying the requirements of the IDS and autopilot systems. In both case studies, the verification outcomes are consistent with domain knowledge or system documentation. The vacuous outcomes further reveal unreachable combinations of states and input ranges, such as high-flow attacks from the safe state in the IDS. 
\end{tcolorbox}

\clearpage
\begingroup
\scriptsize
\setlength{\tabcolsep}{2.5pt}
\renewcommand{\arraystretch}{0.92}
\setlength{\LTcapwidth}{\linewidth}

\begin{longtable}{|>{\raggedright\arraybackslash}p{0.82\linewidth}|>{\centering\arraybackslash}p{0.08\linewidth}|}

\caption{Verification results for the autopilot properties derived from requirement~$\varphi_3$ (Section~\ref{subsec:CaseStudies}). The table reports whether each property is satisfied (\checkmark), violated ($\times$), or vacuously satisfied (v) for state machines learned from the \texttt{Ascent} learning set. Colours indicate the input abstractions and requirement states: \Low{}, \Med{}, and \High{} denote low, medium, and high input ranges, respectively, while \textcolor{unsatred}{red}, \textcolor{satblue}{blue}, and \textcolor{robustgreen}{green} denote states in which requirement~$\varphi_3$ is unsatisfied, satisfied, and robustly satisfied, respectively.}
\label{tab:ap-self-loop-results}
\\

\hline
\rowcolor{gray!20}
\multicolumn{1}{|c|}{\textbf{Property}} & \textbf{Result} \\
\hline
\endfirsthead

\hline
\rowcolor{gray!20}
\multicolumn{1}{|c|}{\textbf{Property}} & \textbf{Result} \\
\hline
\endhead

\hline
\endfoot

\hline
\endlastfoot

\rowcolor{gray!12}
\multicolumn{2}{|c|}{\textbf{With the PitchWheel input}} \\
\hline

From any state in which requirement~$\varphi_3$ is \Unsat, continuously \Low{} PitchWheel input shall eventually cause the system to \Saty\ requirement~$\varphi_3$.
& $\times$ \\
\hline

From any state in which requirement~$\varphi_3$ is \Unsat, continuously \Med{} PitchWheel input shall eventually cause the system to \Saty\ requirement~$\varphi_3$.
& \checkmark \\
\hline

From any state in which requirement~$\varphi_3$ is \Unsat, continuously \High{} PitchWheel input shall eventually cause the system to \Saty\ requirement~$\varphi_3$.
& \checkmark \\
\hline

From any state in which requirement~$\varphi_3$ is \Sat, continuously \Low{} PitchWheel input shall eventually cause the system to \RobustSat\ requirement~$\varphi_3$.
& $\times$ \\
\hline

From any state in which requirement~$\varphi_3$ is \Sat, continuously \Med{} PitchWheel input shall eventually cause the system to \RobustSat\ requirement~$\varphi_3$.
& \checkmark \\
\hline

From any state in which requirement~$\varphi_3$ is \Sat, continuously \High{} PitchWheel input shall eventually cause the system to \RobustSat\ requirement~$\varphi_3$.
& \checkmark \\
\hline

\rowcolor{gray!12}
\multicolumn{2}{|c|}{\textbf{With the Throttle input}} \\
\hline

From any state in which requirement~$\varphi_3$ is \Unsat, continuously \Low{} Throttle input shall eventually cause the system to \Saty\ requirement~$\varphi_3$.
& \checkmark \\
\hline

From any state in which requirement~$\varphi_3$ is \Unsat, continuously \High{} Throttle input shall eventually cause the system to \Saty\ requirement~$\varphi_3$.
& \checkmark \\
\hline

From any state in which requirement~$\varphi_3$ is \Sat, continuously \Low{} Throttle input shall eventually cause the system to \RobustSat\ requirement~$\varphi_3$.
& $\times$ \\
\hline

From any state in which requirement~$\varphi_3$ is \Sat, continuously \High{} Throttle input shall eventually cause the system to \RobustSat\ requirement~$\varphi_3$.
& \checkmark \\
\hline

\rowcolor{gray!12}
\multicolumn{2}{|c|}{\textbf{With both PitchWheel and Throttle inputs}} \\
\hline

From any state in which requirement~$\varphi_3$ is \Unsat, continuously \Low{} PitchWheel and \Low{} Throttle inputs shall eventually cause the system to \Saty\ requirement~$\varphi_3$.
& $\times$ \\
\hline

From any state in which requirement~$\varphi_3$ is \Unsat, continuously \Low{} PitchWheel and \High{} Throttle inputs shall eventually cause the system to \Saty\ requirement~$\varphi_3$.
& v \\
\hline

From any state in which requirement~$\varphi_3$ is \Unsat, continuously \Med{} PitchWheel and \Low{} Throttle inputs shall eventually cause the system to \Saty\ requirement~$\varphi_3$.
& \checkmark \\
\hline

From any state in which requirement~$\varphi_3$ is \Unsat, continuously \Med{} PitchWheel and \High{} Throttle inputs shall eventually cause the system to \Saty\ requirement~$\varphi_3$.
& v \\
\hline

From any state in which requirement~$\varphi_3$ is \Unsat, continuously \High{} PitchWheel and \Low{} Throttle inputs shall eventually cause the system to \Saty\ requirement~$\varphi_3$.
& \checkmark \\
\hline

From any state in which requirement~$\varphi_3$ is \Unsat, continuously \High{} PitchWheel and \High{} Throttle inputs shall eventually cause the system to \Saty\ requirement~$\varphi_3$.
& \checkmark \\
\hline

From any state in which requirement~$\varphi_3$ is \Sat, continuously \Low{} PitchWheel and \Low{} Throttle inputs shall eventually cause the system to \RobustSat\ requirement~$\varphi_3$.
& $\times$ \\
\hline

From any state in which requirement~$\varphi_3$ is \Sat, continuously \Low{} PitchWheel and \High{} Throttle inputs shall eventually cause the system to \RobustSat\ requirement~$\varphi_3$.
& v \\
\hline

From any state in which requirement~$\varphi_3$ is \Sat, continuously \Med{} PitchWheel and \Low{} Throttle inputs shall eventually cause the system to \RobustSat\ requirement~$\varphi_3$.
& \checkmark \\
\hline

From any state in which requirement~$\varphi_3$ is \Sat, continuously \Med{} PitchWheel and \High{} Throttle inputs shall eventually cause the system to \RobustSat\ requirement~$\varphi_3$.
& v \\
\hline

From any state in which requirement~$\varphi_3$ is \Sat, continuously \High{} PitchWheel and \Low{} Throttle inputs shall eventually cause the system to \RobustSat\ requirement~$\varphi_3$.
& $\times$ \\
\hline

From any state in which requirement~$\varphi_3$ is \Sat, continuously \High{} PitchWheel and \High{} Throttle inputs shall eventually cause the system to \RobustSat\ requirement~$\varphi_3$.
& \checkmark \\
\hline

\end{longtable}
\endgroup

\subsection{RQ3: Exploring Unknown Behaviours}
\label{sec:rq3}
We discuss the experiment design and the results obtained for RQ3.

\subsubsection{Experiment Design}
Requirements~$\varphi_1$ and~$\varphi_2$ specify the expected IDS state changes among the \texttt{Safe}, \texttt{Warning}, and \texttt{Alert} states, which are verified in RQ2. Because these requirements do not define the expected behaviour of the \texttt{Tending Warning} (TW) and \texttt{Tending Alert} (TA) states, RQ3 focuses on exploring the behaviour of the IDS in the TW and TA states. To this end, we formulate CTL properties that examine how the IDS evolves from TW and TA under low-, medium-, and high-flow traffic: attack traffic for $\varphi_1$ and normal traffic for $\varphi_2$. Following the verification procedure described in Step~$6$ of \app, we use \textsc{NuSMV} to verify these properties on the five IDS state machines analyzed in RQ2. For the autopilot case study, we examine whether the RQ2 verification results can be used to augment requirement~$\varphi_3$ and identify preconditions for this requirement.

\subsubsection{Results}\mbox{}\\
\textbf{IDS.} Tables~\ref{tab:qc-ctl}(a) and~(b) present the verification results for the IDS properties from the TW and TA states for \texttt{DoS5} and \texttt{DDoS} learning sets, respectively. A property is marked as pass ($\checkmark$) if it holds, fail ($\times$) if it is violated, and vacuous (v) if it holds vacuously. 
Based on Tables~\ref{tab:qc-ctl}(a) and~(b), we observe the following behaviours for the TW and TA states when the IDS is under DoS or DDoS attacks:

\begin{itemize}
    \item The IDS never experiences low-flow DoS attacks while in TW or TA, since all low-flow attack properties for these states are vacuously satisfied.
    \item Medium-flow DoS attacks do not drive the IDS from TW or TA to the highest-criticality state A.
    \item High-flow DoS attacks drive the IDS toward higher criticality states. In particular, from TW, the IDS moves to TA or A; from TA, the IDS either remains in TA or moves to A.
    \item The IDS does not reach the TW or TA states under DDoS attacks.
    
\end{itemize}
   
These observations suggest a refinement of requirement~$\varphi_1$ to capture the observed behaviour of the IDS in the TW and TA states: \textit{When DoS attacks happen, the system shall change state in a staged manner from Safe to Warning, from Warning to Alert, and from Tending Warning and Tending Alert toward higher-criticality states. Medium-flow attacks shall not move the IDS from Tending Warning or Tending Alert to Alert, while high-flow attacks shall move the IDS from Tending Warning or Tending Alert to Tending Alert or Alert.}

Further, based on Table~\ref{tab:qc-ctl}(a), all properties related to $\varphi_2$ for TW and TA are vacuously satisfied. This suggests that TW and TA are attack-driven intermediate states that the IDS leaves as the flow level decreases. As verified in RQ2, normal traffic is then processed in lower-criticality states, where the IDS remains low-criticality or continues recovering toward lower criticality.

\begin{table}[!ht]
\centering
\caption{Verification results for properties derived from the IDS requirements $\varphi_1$ and $\varphi_2$, used to explore IDS behaviours in states \texttt{TW} and \texttt{TA}. The properties related to $\varphi_1$ are evaluated under different attack-traffic intensities, while the properties related to $\varphi_2$ are evaluated under different normal-traffic intensities. The table shows whether each property is satisfied (\checkmark), violated ($\times$), or vacuously satisfied (v).}
\label{tab:qc-ctl}

\footnotesize
\setlength{\tabcolsep}{2.2pt}
\renewcommand{\arraystretch}{1.08}
\vspace*{.4cm}
\begin{tabular}{|p{0.48\linewidth}|C{0.065\linewidth}C{0.065\linewidth}C{0.065\linewidth}|C{0.065\linewidth}C{0.065\linewidth}C{0.065\linewidth}|}
\multicolumn{7}{@{}p{\linewidth}@{}}{\textbf{(a) Results for state machines learned from the \texttt{DoS5} learning set, capturing IDS behaviour under DoS attacks.}} \\
\hline

\multicolumn{1}{|c|}{\multirow{2}{*}{\textbf{Property}}} &
\multicolumn{3}{c|}{\textbf{\texttt{config1}}} &
\multicolumn{3}{c|}{\textbf{\texttt{config3}}} \\
\hhline{|~|---|---|}
\multicolumn{1}{|c|}{} &
\multicolumn{1}{>{\columncolor[HTML]{D9EAD3}}C{0.065\linewidth}}{\textbf{Low}} &
\multicolumn{1}{>{\columncolor[HTML]{FCE5CD}}C{0.065\linewidth}}{\textbf{Med}} &
\multicolumn{1}{>{\columncolor[HTML]{EADCF8}}C{0.065\linewidth}|}{\textbf{High}} &
\multicolumn{1}{>{\columncolor[HTML]{D9EAD3}}C{0.065\linewidth}}{\textbf{Low}} &
\multicolumn{1}{>{\columncolor[HTML]{FCE5CD}}C{0.065\linewidth}}{\textbf{Med}} &
\multicolumn{1}{>{\columncolor[HTML]{EADCF8}}C{0.11\linewidth}|}{\textbf{High}} \\
\hline

\multicolumn{1}{|c|}{\cellcolor{gray!15}\textbf{Properties related to $\varphi_1$}} &
\multicolumn{6}{c|}{\cellcolor{gray!15}\textbf{Attack traffic intensity (Low / Med / High)}} \\
\hline

When attacks happen and the system is in state \texttt{TW}, the system shall transition to \texttt{W} or remain in \texttt{TW}.
& v & \checkmark & $\times$
& v & \checkmark & $\times$ \\
\hline

When attacks happen and the system is in state \texttt{TW}, the system shall transition to \texttt{A} or \texttt{TA}.
& v & $\times$ & \checkmark
& v & $\times$ & \checkmark \\
\hline

When attacks happen and the system is in state \texttt{TA}, the system shall transition to \texttt{W} or \texttt{TW}.
& v & \checkmark & $\times$
& v & \checkmark & $\times$ \\
\hline

When attacks happen and the system is in state \texttt{TA}, the system shall transition to \texttt{A} or remain in \texttt{TA}.
& v & $\times$ & \checkmark
& v & $\times$ & \checkmark \\
\hline

\multicolumn{1}{|c|}{\cellcolor{gray!15}\textbf{Properties related to $\varphi_2$}} &
\multicolumn{6}{c|}{\cellcolor{gray!15}\textbf{Normal traffic intensity (Low / Med / High)}} \\
\hline

When attacks are stopped and the system is in state \texttt{TW}, the system shall transition to \texttt{S}.
& v & v & v
& v & v & v \\
\hline

When attacks are stopped and the system is in state \texttt{TW}, the system shall transition to \texttt{W} or remain in \texttt{TW}.
& v & v & v
& v & v & v \\
\hline

When attacks are stopped and the system is in state \texttt{TA}, the system shall transition to \texttt{W} or \texttt{TW}.
& v & v & v
& v & v & v \\
\hline

When attacks are stopped and the system is in state \texttt{TA}, the system shall transition to \texttt{A} or remain in \texttt{TA}.
& v & v & v
& v & v & v \\
\hline
\end{tabular}

\vspace*{.4cm}
\setlength{\tabcolsep}{2pt}

\begin{tabular}{|p{0.47\linewidth}|C{0.045\linewidth}C{0.045\linewidth}C{0.05\linewidth}|C{0.045\linewidth}C{0.045\linewidth}C{0.05\linewidth}|C{0.045\linewidth}C{0.045\linewidth}C{0.05\linewidth}|}
\multicolumn{10}{@{}p{\linewidth}@{}}{\textbf{(b) Results for state machines learned from the \texttt{DDoS} learning set, capturing IDS behaviour under DDoS attacks.}} \\
\hline

\multicolumn{1}{|c|}{\multirow{2}{*}{\textbf{Property}}} &
\multicolumn{3}{c|}{\textbf{\texttt{config1}}} &
\multicolumn{3}{c|}{\textbf{\texttt{config2}}} &
\multicolumn{3}{c|}{\textbf{\texttt{config3}}} \\
\hhline{|~|---|---|---|}
& \cellcolor[HTML]{D9EAD3}\textbf{Low}
& \cellcolor[HTML]{FCE5CD}\textbf{Med}
& \cellcolor[HTML]{EADCF8}\textbf{High}
& \cellcolor[HTML]{D9EAD3}\textbf{Low}
& \cellcolor[HTML]{FCE5CD}\textbf{Med}
& \cellcolor[HTML]{EADCF8}\textbf{High}
& \cellcolor[HTML]{D9EAD3}\textbf{Low}
& \cellcolor[HTML]{FCE5CD}\textbf{Med}
& \cellcolor[HTML]{EADCF8}\textbf{High} \\
\hline

\multicolumn{1}{|c|}{\cellcolor{gray!15}\textbf{Properties related to $\varphi_1$}} &
\multicolumn{9}{c|}{\cellcolor{gray!15}\textbf{Attack traffic intensity (Low / Med / High)}} \\
\hline

When attacks happen and the system is in state \texttt{TW}, the system shall transition to \texttt{W} or remain in \texttt{TW}.
& v & v & v
& v & v & v
& v & v & v \\
\hline

When attacks happen and the system is in state \texttt{TW}, the system shall transition to \texttt{A} or \texttt{TA}.
& v & v & v
& v & v & v
& v & v & v \\
\hline

When attacks happen and the system is in state \texttt{TA}, the system shall transition to \texttt{W} or \texttt{TW}.
& v & v & v
& v & v & v
& v & v & v \\
\hline

When attacks happen and the system is in state \texttt{TA}, the system shall transition to \texttt{A} or remain in \texttt{TA}.
& v & v & v
& v & v & v
& v & v & v \\
\hline

\multicolumn{1}{|c|}{\cellcolor{gray!15}\textbf{Properties related to $\varphi_2$}} &
\multicolumn{9}{c|}{\cellcolor{gray!15}\textbf{Normal traffic intensity (Low / Med / High)}} \\
\hline

When attacks are stopped and the system is in state \texttt{TW}, the system shall transition to \texttt{S}.
& v & v & v
& v & v & v
& v & v & v \\
\hline

When attacks are stopped and the system is in state \texttt{TW}, the system shall transition to \texttt{W} or remain in \texttt{TW}.
& v & v & v
& v & v & v
& v & v & v \\
\hline

When attacks are stopped and the system is in state \texttt{TA}, the system shall transition to \texttt{W} or \texttt{TW}.
& v & v & v
& v & v & v
& v & v & v \\
\hline

When attacks are stopped and the system is in state \texttt{TA}, the system shall transition to \texttt{A} or remain in \texttt{TA}.
& v & v & v
& v & v & v
& v & v & v \\
\hline
\end{tabular}

\end{table} 
Finally, all verification results in Tables~\ref{tab:qc-ctl}(a) and~(b) are consistent across the state machines learned under \texttt{config1}, \texttt{config2}, and \texttt{config3}.

\textbf{Domain-expert feedback for IDS.}
The explored behaviours matched the domain expert's intuition; nevertheless, the expert found them useful because the learned state machines provide systematic evidence for behaviours that may not be exposed through ad hoc observations or expert experience alone. Further, the expert confirmed that DDoS attacks involve larger numbers of flows than DoS attacks, causing the IDS to bypass the TW and TA states.

\textbf{Autopilot.}
For the autopilot case study, the verification results in RQ2 reveal the preconditions under which requirement~$\varphi_3$ is expected to hold. In particular, reaching the specified altitude within $500$ seconds depends on the aircraft’s nose orientation and throttle. For ascent, the nose should point upward and the throttle should be sufficient. For descent, the nose should point downward. Based on these observations, requirement~$\varphi_3$ can be refined as follows: \textit{when the autopilot is engaged, if the aircraft nose points upward and the throttle is sufficient for ascent, or if the aircraft nose points downward for descent, then the aircraft should reach the specified altitude within $500$ seconds.}

\begin{tcolorbox}[breakable,colback=gray!10!white,colframe=black!75!black]
\textbf{Finding.} Overall, the state machines learned by \app\ support requirement refinement by capturing system behaviours that are not explicitly covered by the original requirements. For the IDS, the DoS results show that the IDS never experiences low-flow attacks or normal traffic while in TW or TA. Medium-flow attacks keep the IDS in warning-related states, while high-flow attacks move the IDS toward alert-related states. Under DDoS attacks, the IDS does not reach TW or TA. For the autopilot, the results reveal the preconditions under which the aircraft is expected to reach the specified altitude.
\end{tcolorbox}

\subsection{Validity Considerations}
\indent\textbf{Internal Validity.} 
To ensure a fair comparison between \app\ and \base, we use identical learning sets and identical test sets across all experiments. The construction of the test sets is not biased toward either technique.
We implemented measures to minimize the impact of extraneous factors. Specifically, for the IDS, (1) we controlled the trace generation process to prevent external traffic not initiated by our simulations from reaching the local users or the router and (2) we monitored the network during the experiments to ensure the absence of anomalies that might have arisen due to events beyond our control. For the autopilot, we kept the simulation setup fixed across runs, including the autopilot Simulink model, controller parameters, and initial conditions, and varied only the input commands used for trace generation.
To construct diverse learning sets, we used adaptive random testing~\citep{metaheuristicsbook} to generate input traces for the IDS and autopilot that explore different regions of the input space while ensuring coverage of all system states. Further, removing inconsistent traces may affect the learned state machines by excluding rare behaviours that are not well captured by the abstraction. To mitigate this threat, \app\ first attempts to resolve inconsistencies by refining the decision-tree-induced intervals and removes traces only when inconsistencies remain after refinement. In our experiments, this step removed only a negligible fraction of the traces ($\sim0.1\%$). In addition, requirement verification requires translating natural-language requirements into CTL formulas. An inaccurate formalization may not fully capture the intended requirement. To mitigate this threat, we derived the CTL formulas using domain knowledge, either in consultation with the domain expert or based on system documentation. This helps ensure that the formal properties reflect the relevant input conditions, system states, and expected successor states. All CTL formulas for the IDS and autopilot are available in our supplementary material~\citep{MELARepoEvaluation}. State-machine generalization during passive learning may produce verification violations that are not supported by the observed executions. We mitigate this threat by checking each violation against the PTA, which captures exactly the observed abstract traces, and by reporting failures in the RQ2 and RQ3 tables only when the violation is present in the PTA.

\textbf{Conclusion Validity.}
We apply the Benjamini-Hochberg (BH) procedure~\citep{benjaminihochberg} to control the false discovery rate. In RQ1, all statistical significance tests were reported using the BH-adjusted p-values. 

\textbf{External Validity.} 
Our experiments are based on two case-study systems: an IDS and an aircraft autopilot system. We selected these systems because they are numeric CPS from two distinct domains, namely network security and avionics. Our IDS case study represents an industrial system for which we could interact with a domain expert~\citep{neginconf}. The aircraft autopilot system is from the Lockheed Martin benchmark and has been previously used in the literature on testing CPS models~\citep{GIANNAKOPOULOU2021106590, TOSEM, nejatievaluating, TSE}. Together, these systems strengthen our evaluation by showing that \app\ can effectively be applied to numeric CPS with different behaviours, inputs, and system characteristics.
Further experiments with a broader range of CPS would strengthen generalizability.

 \section{Related Work}
\label{sec:relwork}
This section compares our work with the relevant strands in two areas: (a)  behavioural model learning, and (b) supervised rule mining for numeric systems. 
\subsection{Behavioural Model Learning}

To better position our work within the literature on learning behavioural models from execution data, we present Table~\ref{tab:relwork_mela} as a structured comparison with the research strands most closely related to~\app. We use the following criteria for the comparison:

\begin{sidewaystable}[p]
\caption{Comparison of~\app~with related work strands on synthesizing behavioural models.}
\label{tab:relwork_mela}
\scriptsize
\setlength{\tabcolsep}{1.2pt} 
\renewcommand{\arraystretch}{1.18} 
\centering 
\begin{tabular}{| L{0.1\textheight}| L{0.1\textheight}| L{0.08\textheight}| L{0.055\textheight}| L{0.110\textheight}| L{0.090\textheight}| L{0.090\textheight}| L{0.125\textheight}| L{0.140\textheight}|}
\hline
\rowcolor{gray!15}
\textbf{Study} &
\textbf{Application Domain} &
\textbf{Learning Paradigm} &
\textbf{Access level} &
\textbf{Learned Model Type} &
\textbf{SUL Input Type} &
\textbf{SUL Output Type} &
\textbf{Transition-label Abstraction} &
\textbf{State-space Abstraction} \\
\hline

CASTLE \cite{TapplerMAK24} &
Stochastic control environments &
Passive \& Active &
Black box &
Markov decision process &
Categorical &
Numeric &
N/A$^{1}$ &
Automatic: dimensionality reduction and k-means clustering \\
\hline

~\cite{AichernigB0HPRR19} &
CPS &
Active &
Black box &
Mealy machine &
Numeric &
Numeric \& Categorical &
Manual &
Manual \\
\hline

FaMoS \cite{PlambeckBHF24} &
CPS &
Passive &
Black box &
Hybrid decision tree &
Numeric &
Numeric &
Automatic: decision tree &
Automatic: dynamic time warping-based clustering \\
\hline

\cite{AichernigKMPT24} &
Reactive systems and protocols &
Passive &
Black box &
Mealy machine &
Categorical &
Categorical &
N/A$^{1}$ &
Automatic: neural networks and hidden-state clustering \\
\hline

GK-tail~\cite{LorenzoliMP08} &
Software interaction &
Passive &
Gray box &
Extended finite state machine &
Numeric \& Categorical &
Numeric \& Categorical &
Automatic: invariant detection &
Automatic: trace-prefix state merging \\
\hline

LEAP \cite{MajumdarMR25} &
Timed reactive systems &
Passive &
Black box &
Event-recording automata &
Numeric \& Categorical &
Categorical &
Manual &
Automatic: SMT-based state merging \\
\hline

\textbf{\app\ } &
CPS &
Passive &
Black box &
Moore Machine &
Numeric \& Categorical &
Numeric \& Categorical &
Automatic: information gain and decision tree &
N/A$^{2}$ \\
\hline

\end{tabular}

\noindent
\begin{minipage}{0.95\textheight}
\raggedright
\footnotesize
$^{1}$ N/A indicates that transition-label abstraction is not required, because discrete events or actions already define transition labels.\\
$^{2}$ N/A indicates that state-space abstraction is not required, because the observed system-state outputs define the state space.
\end{minipage}

\end{sidewaystable} 

(1) \emph{Application domain} denotes the specific types of SULs an approach targets. As shown in Table~\ref{tab:relwork_mela}, some approaches are designed for systems with continuous and time-varying behaviours~\citep{AichernigB0HPRR19,PlambeckBHF24}, some target software systems whose behaviour is captured through system interactions~\citep{LorenzoliMP08}, some focus on reactive and protocol-based systems driven by discrete interactions~\citep{AichernigKMPT24}, and some are developed for sequential decision-making environments with stochastic dynamics~\citep{TapplerMAK24}.

(2) \emph{Learning paradigm} captures how an approach obtains the executions used to construct the behavioural model. Passive approaches rely on a fixed dataset of execution traces collected beforehand~\citep{PlambeckBHF24,AichernigKMPT24,LorenzoliMP08,MajumdarMR25}. Active approaches interact with the SUL during learning, for example by issuing membership or output queries and searching for counterexamples to the current model~\citep{AichernigB0HPRR19}. Hybrid approaches start from an initial trace dataset and collect additional executions during learning to refine the model~\citep{TapplerMAK24}.

(3) \emph{Access level} refers to the level of access required by a behavioural model inference approach. Access level can be categorized as white-box, gray-box, or black-box. A white-box approach~\citep{LorenzoliMP08} requires access to the SUL's source code for instrumentation and trace collection. A gray-box approach assumes partial access to the SUL's internal information, such as selected variables, logs, or interfaces, but not full access to the implementation. A black-box approach operates without access to the SUL's internal implementation and relies only on observable inputs and outputs.

(4) \emph{Learned model type} denotes the state-machine formalism that an approach learns to represent the behaviour of the SUL. Different formalisms capture different behavioural aspects. For example, Moore and Mealy machines differ in whether outputs are associated with states or with input-state interactions~\citep{AichernigKMPT24}; extended finite state machines capture behaviour governed by conditions over data values~\citep{LorenzoliMP08}; and  Markov decision processes and event-recording automata, can capture probabilistic, continuous, or timing-related behaviour~\citep{TapplerMAK24,PlambeckBHF24,MajumdarMR25}.

(5) \emph{SUL input/output type} refers to the types of inputs provided to the SUL and the types of outputs observed from it. For CPS, inputs consist of numeric, time-varying signals, and outputs are typically numeric measurements and may include categorical system-state labels~\citep{AichernigB0HPRR19,PlambeckBHF24}. In reactive and protocol-based systems, inputs are often categorical, such as commands, events, or messages, and outputs are observed as categorical events such as responses or acknowledgements~\citep{AichernigKMPT24}.

(6) \emph{Transition-label abstraction} refers to the process of converting raw trace data into abstract labels for transitions in the learned model. This abstraction is needed when traces contain numeric or continuous that cannot be used directly as transition labels. Some approaches use discrete symbols as transition labels~\citep{TapplerMAK24,AichernigKMPT24}, while others derive more expressive labels in the form of transition conditions, such as numeric ranges or logical predicates~\citep{LorenzoliMP08,PlambeckBHF24}. Abstractions can be defined manually, based on domain knowledge~\citep{AichernigB0HPRR19,MajumdarMR25}, or derived automatically during learning, for example by dynamic invariant detection or decision-tree learning~\citep{LorenzoliMP08,PlambeckBHF24}.

(7) \emph{State-space abstraction} refers to whether an approach derives a finite set of abstract states from the raw system traces, and how these states are obtained. Such abstraction is often needed when system behaviour is observed through continuous, high-dimensional representations that do not directly provide discrete states. In these cases, the approach constructs a state-based model by mapping observations or behaviours to abstract states. This mapping may be defined manually using domain knowledge~\citep{AichernigB0HPRR19} or derived automatically during learning, for example by clustering similar observations or behaviours~\citep{TapplerMAK24,PlambeckBHF24,AichernigKMPT24} or by merging compatible states in a prefix-tree structure~\citep{LorenzoliMP08,MajumdarMR25}.

Having set the stage with Table~\ref{tab:relwork_mela}, we now discuss the specific work strands listed in the table. We begin with approaches for learning behavioural models, followed by techniques for abstracting transition labels and state spaces.

\textit{\textbf{Learned behavioural formalisms.}}
Recent research has explored how to infer behavioural models of SULs from execution traces. These approaches differ in the type of model they learn and in the aspects of behaviour they are able to represent.
Aichernig et al.~\citep{AichernigB0HPRR19,AichernigKMPT24} learn Mealy machines, which associate outputs with transitions and thus capture behaviour as observable responses to inputs.
This formalism is particularly suitable for reactive systems and communication protocols, whose behaviour is naturally viewed as a sequence of observable responses to incoming commands or messages. 
CASTLE~\citep{TapplerMAK24} learns a discrete-state Markov decision process (MDP), which represents action-dependent state changes together with probabilistic transition outcomes. This model is appropriate for stochastic control environments, where the SUL's behaviour depends both on the selected actions and on uncertainty in environment's response.
GK-tail~\citep{LorenzoliMP08} learns an extended finite state machine (EFSM), where states and transitions are enriched with conditions over data values. This representation is suitable for software systems whose executions involve method calls, parameters, and return values, and where different behaviours arise under different data conditions.
FaMoS~\citep{PlambeckBHF24} learns a hybrid automaton, in which the system evolves continuously within discrete control modes and switches between modes when numeric conditions are satisfied. This formalism is particularly suitable for CPS, where discrete control logic interacts with continuously changing physical variables.
LEAP~\citep{MajumdarMR25} learns an event-recording automaton (ERA), which represents behaviour in terms of events and timing constraints that record when events occur. This model is suitable for timed reactive systems, where behaviour depends not only on which events occur but also on the temporal relationships among them.

Our approach, \app, learns a Moore machine, in which outputs are associated with states. This formalism fits our CPS case studies because each trace records a categorical system-state label at every time step. As a result, the learned states are interpretable through these labels, while transitions capture how the system moves between labelled states under abstracted numeric input conditions.

\textit{\textbf{Abstraction for transition labels.}}
Recent research has studied how raw execution data can be transformed into abstract representations for labelling or characterizing transitions in learned behavioural models. Such abstraction is needed when trace data contain numeric information that cannot be used directly as finite transition labels in the target model. Existing approaches differ in how this abstraction is obtained. GK-tail~\citep{LorenzoliMP08} derives transition abstractions in the form of data constraints. It collects parameter values from method calls and returns, applies Daikon~\citep{ernst2007daikon} to infer invariants over these values, and associates the resulting invariants with EFSM transitions.
FaMoS~\citep{PlambeckBHF24} derives transition abstractions as predicates over numeric features. After segmenting executions and clustering them into hybrid modes, it trains decision trees to identify conditions associated with switches between modes; the learned predicates then characterize transitions between the modes.
Aichernig et al.~\citep{AichernigB0HPRR19} abstract continuous inputs and outputs into symbolic categories, which are defined by domain experts and used directly as transition labels in the learned Mealy machine.
LEAP~\citep{MajumdarMR25}, in contrast, does not infer transition abstractions from data; it assumes symbolic constraints on timed events are provided in advance.

\app\ follows the same goal of transforming raw numeric traces into finite transition labels. It uses a decision tree to identify input ranges, maps these ranges to symbolic categories, and uses the resulting categories as transition labels in Moore-machine learning. This produces data-driven and interpretable abstractions that are strongly associated with the system states.

\textit{\textbf{Abstraction for state space.}}
Several studies derive an abstract state space from numeric, continuous, or latent behavioural representations. CASTLE~\citep{TapplerMAK24} maps high-dimensional continuous features to a lower-dimensional representation and then applies k-means clustering to obtain discrete abstract states. FaMoS~\citep{PlambeckBHF24} constructs abstract states by categorizing numeric traces and clustering signal segments with similar dynamical behaviour using a dynamic-time-warping-based distance measure. Aichernig et al.~\citep{AichernigKMPT24} derive abstract states by training a recurrent neural network (RNN) on execution traces and then clustering the resulting hidden-state vectors, which serve as continuous behavioural representations. GK-tail~\citep{LorenzoliMP08} constructs an abstract state space through state merging, grouping trace prefixes that exhibit similar continuation behaviour. Aichernig et al.~\citep{AichernigB0HPRR19} use Mapper, together with manually defined thresholds on continuous variables, to map continuous observations to a finite representation used as the state space for learning. LEAP~\citep{MajumdarMR25} starts from a prefix-tree automaton over symbolic timed words and applies SMT-guided state merging to obtain a compact abstract state space. 

\app\ uses system-state outputs to define the state space when such outputs are available. These outputs also guide the decision-tree-based abstraction of numeric inputs and become the outputs of Moore-machine states. \app\ is not limited to this source of state information: it can use abstract states derived through machine-learning-based abstraction or state merging. In our case studies, the observed system-state outputs already provide a discrete and behaviourally meaningful state space.

\subsection{Supervised Rule Mining}
Supervised rule-mining techniques, such as decision trees and decision rules, are effective for identifying predicates that relate system inputs to system states. Prior work has used these techniques mainly to characterize or explain specific SUL behaviours, for example, by identifying input conditions associated with particular program outcomes, such as passing, failing, or non-robust behaviours~\citep{kampmann2020does,JodatNSS23,TOSEM}. However, these techniques capture static relationships in the data and are not designed to model temporal relationships between states or input-driven state changes over time. In contrast, \app\ uses rule mining to derive interpretable input partitions for numeric traces and then applies passive automata learning to infer the temporal structure of the SUL's behaviour. This combination preserves the interpretability of rule mining while addressing its limitation in modelling state-based temporal behaviour.

 \section{Conclusion}
\label{sec:con}
In this article, we presented \app, a passive automata learning approach enhanced with statistical machine learning for synthesizing behavioural models of CPS from numeric time-series data. \app\ learns interpretable state machines that can be used to check whether system requirements hold and to examine behaviours that the requirements do not fully specify. Across the network intrusion detection and autopilot case studies, the verification and exploration results were consistent with domain expertise, showing that the learned models capture meaningful system behaviour. Future work will explore how learned state machines can be adapted into runtime monitors for CPS, so that deviations from learned behaviour can be detected online.

\noindent\textbf{Acknowledgements}  We gratefully acknowledge the financial support received from  RabbitRun Technologies Inc, Mitacs, and NSERC of Canada through the Discovery and I2I programs.

\noindent\textbf{Author Contributions} Negin Ayoughi, Baharin A. Jodat, Shiva Nejati, and Mehrdad Sabetzadeh contributed to the conceptualization and methodology of the study. Negin Ayoughi, Baharin A. Jodat, and Armina Faghihi conducted the formal analysis and investigation. Negin Ayoughi and Baharin A. Jodat prepared the original draft. All authors contributed to reviewing and editing the manuscript. Patricio Saavedra, Shiva Nejati, and Mehrdad Sabetzadeh provided resources. Shiva Nejati and Mehrdad Sabetzadeh acquired funding and supervised the study.

\noindent\textbf{Data Availability} We publicly share the implementation of \app\ in our online repository~\citep{MELARepo}, together with the experimental data and learned models used in our evaluation. The repository is organized by case study into five categories. \texttt{Code} provides the scripts for trace creation, input-variable selection, range abstraction with \app\ and \base, and automata learning~\citep{MELARepoCode}. \texttt{Data} includes the generated time-series data, the derived learning sets, and the abstracted traces used as input to automata learning~\citep{MELARepoData}. \texttt{Evaluation} provides the scripts and test sets for RQ1, RQ2, and RQ3, including conformance checking and \textsc{NuSMV} model checking~\citep{MELARepoEvaluation}. \texttt{Result} reports the learned Moore machines, PTAs, the RQ1 comparison between \app\ and \base, and the RQ2 and RQ3 results on requirement verification and behaviour exploration~\citep{MELARepoResult}. \texttt{Testbed} provides the material needed to reproduce the case-study workflows, including the IDS traffic-generation scripts and the Simulink autopilot benchmark~\citep{MELARepoTestbed}.

\subsection*{Declarations}
\textbf{Competing Interests} The authors declare no competing interests.

\bibliography{bibliography}

@misc{metasploit,
  title = {Metasploit Framework},
  author = {{Rapid7}},
  year = {2024},
  howpublished= {\url{https://www.metasploit.com/}},
  note = {Last accessed: June 2026},
}

@inproceedings{muvskardin2022aalpy,
  author       = {Edi Muskardin and
                  Bernhard K. Aichernig and
                  Ingo Pill and
                  Andrea Pferscher and
                  Martin Tappler},
  Xeditor       = {Zhe Hou and
                  Vijay Ganesh},
  title        = {AALpy: An Active Automata Learning Library},
  booktitle    = {Proceedings of 19th International Symposium on Automated Technology for Verification and Analysis ({ATVA} 2021)},
  Xseries       = {Lecture Notes in Computer Science},
  volume       = {12971},
  pages        = {67--73},
  publisher    = {Springer},
  year         = {2021},
  address      = {Berlin, Germany},
  doi          = {10.1007/S11334-022-00449-3}
}

@inproceedings{Hall11,
  author       = {Mathew Hall},
  Xeditor       = {Myra B. Cohen and
                  Mel {\'{O}} Cinn{\'{e}}ide},
  title        = {Complexity Metrics for Hierarchical State Machines},
  booktitle    = {Proceedings of Third International Symposium on Search Based Software Engineering ({SSBSE} 2011)},
  Xseries       = {Lecture Notes in Computer Science},
  volume       = {6956},
  pages        = {76--81},
  publisher    = {Springer},
  year         = {2011},
  address      = {Berlin, Germany},
  doi = {10.1007/978-3-642-23716-4_10}
}

@inproceedings{BeerBER97,
  author       = {Ilan Beer and
                  Shoham Ben{-}David and
                  Cindy Eisner and
                  Yoav Rodeh},
  Xeditor       = {Orna Grumberg},
  title        = {Efficient Detection of Vacuity in {ACTL} Formulaas},
  booktitle    = {Proceedings of 9th International Conference on Computer Aided Verification ({CAV} 1997)},
  Xseries       = {Lecture Notes in Computer Science},
  volume       = {1254},
  pages        = {279--290},
  publisher    = {Springer},
  year         = {1997},
  address      = {Berlin, Germany},
  doi          = {10.1007/3-540-63166-6\_28}
}

@book{mcbook,
  author       = {Edmund M. Clarke and
                  Orna Grumberg and
                  Daniel Kroening and
                  Doron A. Peled and
                  Helmut Veith},
  title        = {Model checking, 2nd Edition},
  publisher    = {{MIT} Press},
  year         = {2018},
  address      = {Cambridge, MA, USA},
  note={{I}SBN: 9780262038836} 
}

@inproceedings{Pnueli77,
  author       = {Amir Pnueli},
  title        = {The Temporal Logic of Programs},
  booktitle    = {Proceedings of 18th Annual Symposium on Foundations of Computer Science},
  pages        = {46--57},
  publisher    = {{IEEE} Computer Society},
  year         = {1977},
  address      = {Piscataway, NJ, USA},
  doi          = {10.1109/SFCS.1977.32}
}

@inproceedings{muskardin2022active,
  author       = {Bernhard K. Aichernig and
                  Edi Muskardin and
                  Andrea Pferscher},
  Xeditor       = {Matt Luckcuck and
                  Marie Farrell},
  title        = {Active vs. Passive: {A} Comparison of Automata Learning Paradigms
                  for Network Protocols},
  booktitle    = {Proceedings of Fourth International Workshop on Formal Methods for Autonomous Systems and Fourth International Workshop on Automated and verifiable Software sYstem DEvelopment ({FMAS/ASYDE} 2022)},
  Xseries       = {{EPTCS}},
  volume       = {371},
  pages        = {1--19},
  year         = {2022},
  publisher    = {Electronic Proceedings in Theoretical Computer Science (EPTCS)},
  address      = {Open Access Platform},
  doi          = {10.4204/EPTCS.371.1}
}

@article{PferscherA22,
  author       = {Andrea Pferscher and
                  Bernhard K. Aichernig},
  title        = {Fingerprinting and analysis of Bluetooth devices with automata learning},
  journal      = {Formal Methods Syst. Des.},
  volume       = {61},
  number       = {1},
  pages        = {35--62},
  year         = {2022},
  doi          = {10.1007/S10703-023-00425-Y}
}

@inproceedings{tappler2017model,
  title={Model-based testing IoT communication via active automata learning},
  author={Tappler, Martin and Aichernig, Bernhard K and Bloem, Roderick},
  booktitle={2017 IEEE International conference on software testing, verification and validation (ICST)},
  pages={276--287},
  year={2017},
  organization={IEEE},
  publisher= {IEEE},
  address= {Piscataway, NJ, USA},
  doi = {10.1109/ICST.2017.32}
}

@book{de2010grammatical,
  author={De la Higuera, Colin},
  title={Grammatical inference: learning automata and grammars},
 publisher={Cambridge University Press},
  year={2010},
   address      = {Cambridge, UK}
}

@inproceedings{cano2010inferring,
  author       = {Antonio Cano G{\'{o}}mez},
  Xeditor       = {Jos{\'{e}} M. Sempere and
                  Pedro Garc{\'{\i}}a},
  title        = {Inferring Regular Trace Languages from Positive and Negative Samples},
  booktitle    = {Proceedings of 10th International Colloquium on Grammatical Inference ({ICGI} 2010)},
  Xseries       = {Lecture Notes in Computer Science},
  volume       = {6339},
  pages        = {11--23},
  publisher    = {Springer},
  year         = {2010},
  address      = {Berlin, Germany},
  doi ={10.1007/978-3-642-15488-1_3}
}

@inproceedings{GarhewalD23,
  author       = {Bharat Garhewal and
                  Carlos Diego Nascimento Damasceno},
  title        = {An Experimental Evaluation of Conformance Testing Techniques in Active
                  Automata Learning},
  booktitle    = {Proceedings of 26th {ACM/IEEE} International Conference on Model Driven Engineering Languages and Systems ({MODELS} 2023)},
  pages        = {217--227},
  publisher    = {{IEEE}},
  year         = {2023},
  address      = {Piscataway, NJ, USA},
  doi          = {10.1109/MODELS58315.2023.00012}
}

@article{Vaandrager17,
  author       = {Frits W. Vaandrager},
  title        = {Model learning},
  journal      = {Commun. {ACM}},
  volume       = {60},
  number       = {2},
  pages        = {86--95},
  year         = {2017},
  doi = {10.1145/2967606}
}

@inproceedings{NeiderSVK97,
  author       = {Daniel Neider and
                  Rick Smetsers and
                  Frits W. Vaandrager and
                  Harco Kuppens},
  Xeditor       = {Tiziana Margaria and
                  Susanne Graf and
                  Kim G. Larsen},
  title        = {Benchmarks for Automata Learning and Conformance Testing},
  booktitle    = {Models, Mindsets, Meta: The What, the How, and the Why Not? - Essays
                  Dedicated to Bernhard Steffen on the Occasion of His 60th Birthday},
  Xseries       = {Lecture Notes in Computer Science},
  volume       = {11200},
  pages        = {390--416},
  publisher    = {Springer},
  year         = {2018},
  address      = {Berlin, Germany},
  doi          = {10.1007/978-3-030-22348-9\_23}
}

@inproceedings{BergA25,
  author       = {Benjamin von Berg and
                  Bernhard K. Aichernig},
  Xeditor       = {Ruzica Piskac and
                  Zvonimir Rakamaric},
  title        = {Extending AALpy with Passive Learning: {A} Generalized State-Merging
                  Approach},
  booktitle    = {Proceedings of 37th International Conference on Computer Aided Verification ({CAV} 2025)},
  Xseries       = {Lecture Notes in Computer Science},
  volume       = {15934},
  pages        = {127--140},
  publisher    = {Springer},
  year         = {2025},
  address   = {Zagreb, Croatia},
  doi       = {10.1007/978-3-031-98685-7_6}
}

@inproceedings{JodatNSS23,
  author       = {Baharin Aliashrafi Jodat and
                  Shiva Nejati and
                  Mehrdad Sabetzadeh and
                  Patricio Saavedra},
  title        = {Learning Non-robustness using Simulation-based Testing: a Network
                  Traffic-shaping Case Study},
  booktitle    = {Proceedings of {IEEE} Conference on Software Testing, Verification and Validation ({ICST} 2023)},
  pages        = {386--397},
  publisher    = {{IEEE}},
  year         = {2023},
  address      = {Piscataway, NJ, USA},
  doi          = {10.1109/ICST57152.2023.00043}
}

@article{pareto18,
  title={Pareto efficient multi-objective black-box test case selection for simulation-based testing},
  author={Arrieta, Aitor and Wang, Shuai and Markiegi, Urtzi and Arruabarrena, Ainhoa and Etxeberria, Leire and Sagardui, Goiuria},
  journal={Information and Software Technology},
  volume={114},
  pages={137--154},
  year={2019},
  publisher={Elsevier},
  doi={10.1016/j.infsof.2019.06.009}
  }

@article{signals19,
  title={Requirements-driven test generation for autonomous vehicles with machine learning components},
  author={Tuncali, Cumhur Erkan and Fainekos, Georgios and Prokhorov, Danil and Ito, Hisahiro and Kapinski, James},
  journal={IEEE Transactions on Intelligent Vehicles},
  volume={5},
  number={2},
  pages={265--280},
  year={2019},
  publisher={IEEE},
  doi = {10.1109/TIV.2019.2955903}
}

@book{offuttTesting,
  author       = {Paul Ammann and
                  Jeff Offutt},
  title        = {Introduction to Software Testing},
  publisher    = {Cambridge University Press},
  year         = {2008},
  address      = {Cambridge, UK}
}

@article{TOSEM,
  author = {Jodat, Baharin A. and Chandar, Abhishek and Nejati, Shiva and Sabetzadeh, Mehrdad},
title = {Test Generation Strategies for Building Failure Models and Explaining Spurious Failures},
year = {2024},
issue_date = {May 2024},
publisher = {Association for Computing Machinery},
address = {New York, NY, USA},
volume = {33},
number = {4},
issn = {1049-331X},
url = {https://doi.org/10.1145/3638246},
doi = {10.1145/3638246},
journal = {ACM Trans. Softw. Eng. Methodol.},
month = apr,
articleno = {93},
numpages = {32}
}

@article{zargar2013survey,
  author       = {Saman Taghavi Zargar and
                  James Joshi and
                  David Tipper},
  title        = {A Survey of Defense Mechanisms Against Distributed Denial of Service
                  (DDoS) Flooding Attacks},
  journal      = {{IEEE} Commun. Surv. Tutorials},
  volume       = {15},
  number       = {4},
  pages        = {2046--2069},
  year         = {2013},
  doi = {10.1109/SURV.2013.031413.00127}
}

@inproceedings{fiteruau2016combining,
  author       = {Paul Fiterau{-}Brostean and
                  Ramon Janssen and
                  Frits W. Vaandrager},
  Xeditor       = {Swarat Chaudhuri and
                  Azadeh Farzan},
  title        = {Combining Model Learning and Model Checking to Analyze {TCP} Implementations},
  booktitle    = {Proceedings of 28th International Conference on Computer Aided Verification ({CAV} 2016)},
  Xseries       = {Lecture Notes in Computer Science},
  volume       = {9780},
  pages        = {454--471},
  publisher    = {Springer},
  year         = {2016},
  address      = {Berlin, Germany},
  doi= {10.1007/978-3-319-41540-6_25}
}

@misc{MELARepo,
key      = {CPS-Behavioural-Model-Synthesis},
  author = {Ayoughi, Negin and others},
  title = {{Artifacts and Supplementary Material}},
  year = {2026},
  howpublished = {GitHub repository},
  note  = {Artifacts and Supplementary Material [Online]. Available: \url{https://github.com/neayoughi/CPS-Behavioural-Model-Synthesis.git}}
}

@misc{MELARepoCode,
key      = {CPS-Behavioural-Model-Synthesis},
  author = {Ayoughi, Negin and others},
  title = {{Code for CPS Behavioural Model Synthesis with MELA}},
  year = {2026},
  howpublished = {GitHub repository},
  note  = {Artifacts and Supplementary Material [Online]. Available: \url{https://github.com/neayoughi/CPS-Behavioural-Model-Synthesis/tree/autopilot-final-mela-artifacts/Code}}
}

@misc{MELARepoData,
key      = {CPS-Behavioural-Model-Synthesis},
  author = {Ayoughi, Negin and others},
  title = {{Data for CPS Behavioural Model Synthesis with MELA}},
  year = {2026},
  howpublished = {GitHub repository},
  note  = {Artifacts and Supplementary Material [Online]. Available: \url{https://github.com/neayoughi/CPS-Behavioural-Model-Synthesis/tree/update-combined-readme/Data}}
}

@misc{MELARepoEvaluation,
key      = {CPS-Behavioural-Model-Synthesis},
  author = {Ayoughi, Negin and others},
  title = {{Evaluation Material for CPS Behavioural Model Synthesis with MELA}},
  year = {2026},
  howpublished = {GitHub repository},
  note  = {Artifacts and Supplementary Material [Online]. Available: \url{https://github.com/neayoughi/CPS-Behavioural-Model-Synthesis/tree/autopilot-final-mela-artifacts/Evalution}}
}

@misc{MELARepoResult,
key      = {CPS-Behavioural-Model-Synthesis},
  author = {Ayoughi, Negin and others},
  title = {{Results for CPS Behavioural Model Synthesis with MELA}},
  year = {2026},
  howpublished = {GitHub repository},
  note  = {Artifacts and Supplementary Material [Online]. Available: \url{https://github.com/neayoughi/CPS-Behavioural-Model-Synthesis/tree/autopilot-final-mela-artifacts/Results}}
}

@misc{descentrq2,
key      = {CPS-Behavioural-Model-Synthesis},
  author = {Ayoughi, Negin and others},
  title = {{RQ2 results for the autopilot case study using the \texttt{Descent} learning set}},
  year = {2026},
  howpublished = {GitHub repository},
  note  = {Artifacts and Supplementary Material [Online]. Available: \url{https://github.com/neayoughi/CPS-Behavioural-Model-Synthesis/tree/update-combined-readme/Results/Autopilot/RQ2/summary/Descend_RQ2.pdf}}
}

@misc{MELARepoTestbed,
key      = {CPS-Behavioural-Model-Synthesis},
  author = {Ayoughi, Negin and others},
  title = {{Testbeds for CPS Behavioural Model Synthesis with MELA}},
  year = {2026},
  howpublished = {GitHub repository},
  note  = {Artifacts and Supplementary Material [Online]. Available: \url{https://github.com/neayoughi/CPS-Behavioural-Model-Synthesis/tree/autopilot-final-mela-artifacts/Testbed}}
}

@misc{lockheedmartin,
author       = {{Lockheed Martin Corporation}},
  title = {Lockheed Martin},
  year = {2026},  howpublished= {\url{https://www.lockheedmartin.com}},
  note        = {Last accessed: June 2026},
}

@misc{clutchlockup,
 author = {{The MathWorks, Inc.}},
 title = {Building a Clutch Lock-Up Model},
 year        = {2026},
 howpublished= {\url{https://www.mathworks.com/help/simulink/slref/building-a-clutch-lock-up-model.html}},
 note        = {Last accessed: June 2026},
}

@inproceedings{khandait2024arch,
  title={ARCH-COMP 2024 Category Report: Falsification},
  author={Tanmay Khandait and Federico Formica and Paolo Arcaini and Surdeep Chotaliya and Georgios Fainekos and Abdelrahman Hekal and Atanu Kundu and Ethan Lew and Michele Loreti and Claudio Menghi and Laura Nenzi and Giulia Pedrielli and Jarkko Peltomäki and Ivan Porres and Rajarshi Ray and Valentin Soloviev and Ennio Visconti and Masaki Waga and Zhenya Zhang},
  booktitle={Proceedings of the 11th Int. Workshop on Applied Verification for Continuous and Hybrid Systems},
  volume={103},
  pages={122--144},
  year={2024},
  doi         = {10.29007/hgfv},
  series    = {EPiC Series in Computing},
  publisher = {EasyChair},
  address   = {Boulder, CO, USA}
}

@misc{cruisecontroller,
author = {{The MathWorks, Inc.}},
title = {Cruise Control Test Generation},
  year        = {2026},
  howpublished= {\url{https://www.mathworks.com/help/sldv/ug/cruise-control-test-generation.html}},
  note        = {Last accessed: June 2026},
}

@misc{guidancecontrol,
author = {{The MathWorks, Inc.}},
  title       = {Design a Guidance System in {MATLAB} and {Simulink}},
  year        = {2026},
  howpublished= {\url{https://www.mathworks.com/help/simulink/slref/designing-a-guidance-system-in-matlab-and-simulink.html}},
  note        = {Last accessed: June 2026},
}

@misc{dcmotor,
author = {{Sam Elshamy}},
title = {DC Motor Model Simulink Model},
  year        = {2026},
  howpublished= {\url{https://www.mathworks.com/matlabcentral/fileexchange/11587-dc-motor-model-simulink}},
  note        = {Last accessed: June 2026},
}

@inproceedings{neginconf,
  title={Enhancing automata learning with statistical machine learning: A network security case study},
  author={Ayoughi, Negin and Nejati, Shiva and Sabetzadeh, Mehrdad and Saavedra, Patricio},
  booktitle={Proceedings of the ACM/IEEE 27th International Conference on Model Driven Engineering Languages and Systems},
  pages={172--182},
  year={2024},
  address = {Linz, Austria},
  doi = {10.1145/3640310.3674087},
  publisher = {Association for Computing Machinery}
}

@article{vargha2000critique,
  title={A critique and improvement of the CL common language effect size statistics of McGraw and Wong},
  author={Vargha, Andr{\'a}s and Delaney, Harold},
  journal={Journal of Educational and Behavioral Statistics},
  volume={25},
  number={2},
  pages={101--132},
  year={2000},
  publisher={Sage Publications Sage CA: Los Angeles, CA},
  doi       = {10.3102/10769986025002101}
}

@book{Mann-Whitney,
  title     = {Practical Nonparametric Statistics},
  author    = {Conover, William},
  edition   = {3},
  year      = {1999},
  publisher = {Wiley \& Sons},
  address   = {New York}
}

@article{TSE,
  author       = {Baharin Aliashrafi Jodat and
                  Khouloud Gaaloul and
                  Mehrdad Sabetzadeh and
                  Shiva Nejati},
  title        = {Automated Test Validators for Flaky Cyber-Physical System Simulators:
                  Approach and Evaluation},
  journal      = {IEEE Transactions on Software Engineering},
  volume       = {abs/2508.20902},
  year         = {2026},
  url          = {https://doi.org/10.1109/TSE.2026.3685556},
  doi          = {10.1109/TSE.2026.3685556}
}

@book{autopilothandbook,
  author    = {{Federal Aviation Administration}},
  title     = {Advanced Avionics Handbook},
  year      = {2009},
  publisher = {Aviation Supplies \& Academics},
  address   = {Newcastle, WA, USA},
  series    = {FAA Handbooks Series},
  isbn      = {9781560277583},
  url       = {https://books.google.lu/books?id=2xGuPwAACAAJ}
}

@book{federal2009pilot,
  title={Pilot's handbook of aeronautical knowledge},
  author={{Federal Aviation Administration}},
  year={2009},
  publisher={Skyhorse Publishing Inc.},
  address = {New York, NY, USA}
}

@article{GIANNAKOPOULOU2021106590,
title = {Automated formalization of structured natural language requirements},
journal = {Information and Software Technology},
volume = {137},
pages = {106590},
year = {2021},
issn = {0950-5849},
doi = {10.1016/j.infsof.2021.106590},
Xurl = {https://www.sciencedirect.com/science/article/pii/S0950584921000707},
author = {Dimitra Giannakopoulou and Thomas Pressburger and Anastasia Mavridou and Johann Schumann}
}

@inproceedings{nejatievaluating,
author = {Nejati, Shiva and Gaaloul, Khouloud and Menghi, Claudio and Briand, Lionel C. and Foster, Stephen and Wolfe, David},
title = {Evaluating model testing and model checking for finding requirements violations in Simulink models},
year = {2019},
isbn = {9781450355728},
publisher = {Association for Computing Machinery},
address = {New York, NY, USA},
Xurl = {https://doi.org/10.1145/3338906.3340444},
doi = {10.1145/3338906.3340444},
booktitle = {Proceedings of 27th {ACM} Joint Meeting on European Software Engineering Conference and Symposium on the Foundations of Software Engineering ({ESEC/FSE} 2019)},
pages = {1015–1025},
numpages = {11},
Xlocation = {Tallinn, Estonia},
Xseries = {ESEC/FSE 2019}
}

@article{benjaminihochberg,
    author = {Benjamini, Yoav and Hochberg, Yosef},
    title = {Controlling the False Discovery Rate: A Practical and Powerful Approach to Multiple Testing},
    journal = {Journal of the Royal Statistical Society: Series B (Methodological)},
    volume = {57},
    number = {1},
    pages = {289-300},
    year = {2018},
    issn = {0035-9246},
    doi = {10.1111/j.2517-6161.1995.tb02031.x},
    Xurl = {https://doi.org/10.1111/j.2517-6161.1995.tb02031.x},
    Xeprint = {https://academic.oup.com/jrsssb/article-pdf/57/1/289/49173396/jrsssb_57_1_289.pdf},
}

@inproceedings{ovsiannikova2018active,
  title={Active learning of formal plant models for cyber-physical systems},
  author={Ovsiannikova, Polina and Chivilikhin, Daniil and Ulyantsev, Vladimir and Stankevich, Andrey and Zakirzyanov, Ilya and Vyatkin, Valeriy and Shalyto, Anatoly},
  booktitle={2018 IEEE 16th International Conference on Industrial Informatics (INDIN)},
  pages={719--724},
  year={2018},
  publisher = {IEEE},
  address   = {Porto, Portugal},
  doi          = {10.1109/INDIN.2018.8471924}
}

@inproceedings{aarts2010inference,
  title={Inference and abstraction of the biometric passport},
  author={Aarts, Fides and Schmaltz, Julien and Vaandrager, Frits},
  booktitle={International Symposium On Leveraging Applications of Formal Methods, Verification and Validation},
  pages={673--686},
  year={2010},
  publisher = {Springer},
  address = {Heraklion, Crete, Greece},
  doi = {10.1007/978-3-642-16558-0_54}
}

@inproceedings{hajnorouzi2025model,
  author       = {Mehrnoush Hajnorouzi and
                  Astrid Rakow and
                  Martin Fr{\"{a}}nzle},
  Xeditor       = {Matt Luckcuck and
                  Maike Schwammberger and
                  Mengwei Xu},
  title        = {Model Learning for Adjusting the Level of Automation in {HCPS}},
  booktitle    = {Proceedings of 7th International Workshop on Formal Methods for Autonomous Systems ({FMAS@iFM} 2025)},
  series       = {{EPTCS}},
  pages        = {96--113},
  year         = {2025},
  Xmonth        = nov,
  Xurl          = {https://doi.org/10.4204/EPTCS.436.10},
  doi          = {10.4204/EPTCS.436.10},
  Xtimestamp    = {Tue, 24 Mar 2026 08:44:57 +0100},
  Xbiburl       = {https://dblp.org/rec/journals/corr/abs-2511-14437.bib},
  Xbibsource    = {dblp computer science bibliography, https://dblp.org},
  publisher = {Open Publishing Association},
  address = {Paris, France}
}

@article{GiantamidisTB21,
  author       = {Georgios Giantamidis and
                  Stavros Tripakis and
                  Stylianos Basagiannis},
  title        = {Learning Moore machines from input-output traces},
  journal      = {Int. J. Softw. Tools Technol. Transf.},
  volume       = {23},
  number       = {1},
  pages        = {1--29},
  year         = {2021},
  doi          = {10.1007/S10009-019-00544-0}
}

@inproceedings{TapplerMAK24,
  author       = {Martin Tappler and
                  Edi Muskardin and
                  Bernhard K. Aichernig and
                  Bettina K{\"{o}}nighofer},
  title        = {Learning Environment Models with Continuous Stochastic Dynamics -
                  with an Application to Deep {RL} Testing},
  booktitle    = {Proceedings of {IEEE} Conference on Software Testing, Verification and Validation ({ICST} 2024)},
  pages        = {197--208},
  publisher    = {{IEEE}},
  address   = {Toronto, ON, Canada},
  year         = {2024},
  Xurl          = {https://doi.org/10.1109/ICST60714.2024.00026},
  doi          = {10.1109/ICST60714.2024.00026},
  Xtimestamp    = {Fri, 04 Jul 2025 22:07:57 +0200},
  Xbiburl       = {https://dblp.org/rec/conf/icst/TapplerMAK24.bib},
  Xbibsource    = {dblp computer science bibliography, https://dblp.org}
}

@inproceedings{AichernigB0HPRR19,
  author       = {Bernhard K. Aichernig and
                  Roderick Bloem and
                  Masoud Ebrahimi and
                  Martin Horn and
                  Franz Pernkopf and
                  Wolfgang Roth and
                  Astrid Rupp and
                  Martin Tappler and
                  Markus Tranninger},
  Xeditor       = {Christophe Gaston and
                  Nikolai Kosmatov and
                  Pascale Le Gall},
  title        = {Learning a Behavior Model of Hybrid Systems Through Combining Model-Based
                  Testing and Machine Learning},
  booktitle    = {Proceedings of 31st {IFIP} {WG} 6.1 International Conference on Testing Software and Systems ({ICTSS} 2019)},
  Xseries       = {Lecture Notes in Computer Science},
  volume       = {11812},
  pages        = {3--21},
  publisher    = {Springer},
  address = {Paris, France},
  year         = {2019},
  Xurl          = {https://doi.org/10.1007/978-3-030-31280-0\_1},
  doi          = {10.1007/978-3-030-31280-0\_1},
  Xtimestamp    = {Mon, 03 Mar 2025 21:20:13 +0100},
  Xbiburl       = {https://dblp.org/rec/conf/pts/AichernigB0HPRR19.bib},
  Xbibsource    = {dblp computer science bibliography, https://dblp.org}
}

@inproceedings{PlambeckBHF24,
  author       = {Swantje Plambeck and
                  Aaron Bracht and
                  Nemanja Hranisavljevic and
                  G{\"{o}}rschwin Fey},
  Xeditor       = {Erika {\'{A}}brah{\'{a}}m and
                  Manuel Mazo Jr.},
  title        = {FaMoS- Fast Model Learning for Hybrid Cyber-Physical Systems using
                  Decision Trees},
  booktitle    = {Proceedings of 27th {ACM} International Conference on Hybrid Systems: Computation and Control ({HSCC} 2024)},
  pages        = {7:1--7:10},
  publisher = {Association for Computing Machinery (ACM)},
  address   = {Hong Kong SAR, China},
  year         = {2024},
  Xurl          = {https://doi.org/10.1145/3641513.3650131},
  doi          = {10.1145/3641513.3650131},
  Xtimestamp    = {Tue, 02 Sep 2025 10:14:11 +0200},
  Xbiburl       = {https://dblp.org/rec/conf/hybrid/PlambeckBHF24.bib},
  Xbibsource    = {dblp computer science bibliography, https://dblp.org}
}

@article{AichernigKMPT24,
  author       = {Bernhard K. Aichernig and
                  Sandra K{\"{o}}nig and
                  Cristinel Mateis and
                  Andrea Pferscher and
                  Martin Tappler},
  title        = {Learning minimal automata with recurrent neural networks},
  journal      = {Softw. Syst. Model.},
  volume       = {23},
  number       = {3},
  pages        = {625--655},
  year         = {2024},
  Xurl          = {https://doi.org/10.1007/s10270-024-01160-6},
  doi          = {10.1007/S10270-024-01160-6},
  Xtimestamp    = {Tue, 01 Apr 2025 19:04:02 +0200},
  Xbiburl       = {https://dblp.org/rec/journals/sosym/AichernigKMPT24.bib},
  Xbibsource    = {dblp computer science bibliography, https://dblp.org}
}

@inproceedings{LorenzoliMP08,
  author       = {Davide Lorenzoli and
                  Leonardo Mariani and
                  Mauro Pezz{\`{e}}},
  Xeditor       = {Wilhelm Sch{\"{a}}fer and
                  Matthew B. Dwyer and
                  Volker Gruhn},
  title        = {Automatic generation of software behavioral models},
  booktitle    = {Proceedings of 30th International Conference on Software Engineering ({ICSE} 2008)},
  pages        = {501--510},
  publisher = {Association for Computing Machinery (ACM)},
  year         = {2008},
  Xurl          = {https://doi.org/10.1145/1368088.1368157},
  address   = {Leipzig, Germany},
  doi          = {10.1145/1368088.1368157},
  Xtimestamp    = {Sun, 02 Oct 2022 16:06:31 +0200},
  Xbiburl       = {https://dblp.org/rec/conf/icse/LorenzoliMP08.bib},
  Xbibsource    = {dblp computer science bibliography, https://dblp.org}
}

@inproceedings{MajumdarMR25,
  author       = {Anirban Majumdar and
                  Sayan Mukherjee and
                  Jean{-}Fran{\c{c}}ois Raskin},
  Xeditor       = {Meenakshi D'Souza and
                  Raghavan Komondoor and
                  B. Srivathsan},
  title        = {Learning Event-Recording Automata Passively},
  booktitle    = {Proceedings of 23rd International Symposium on Automated Technology for Verification and Analysis ({ATVA} 2025)},
  Xseries       = {Lecture Notes in Computer Science},
  volume       = {16145},
  pages        = {27--48},
  publisher    = {Springer},
  year         = {2025},
  Xurl          = {https://doi.org/10.1007/978-3-032-08707-2\_2},
  address   = {Bengaluru, India},
  doi       = {10.1007/978-3-032-08707-2_2},
  Xtimestamp    = {Sat, 15 Nov 2025 13:44:41 +0100},
  Xbiburl       = {https://dblp.org/rec/conf/atva/MajumdarMR25.bib},
  Xbibsource    = {dblp computer science bibliography, https://dblp.org}
}

@inproceedings{CimattiCGGPRST02,
  author       = {Alessandro Cimatti and
                  Edmund M. Clarke and
                  Enrico Giunchiglia and
                  Fausto Giunchiglia and
                  Marco Pistore and
                  Marco Roveri and
                  Roberto Sebastiani and
                  Armando Tacchella},
  Xeditor       = {Ed Brinksma and
                  Kim Guldstrand Larsen},
  title        = {NuSMV 2: An OpenSource Tool for Symbolic Model Checking},
  booktitle    = {Proceedings of 14th International Conference on Computer Aided Verification ({CAV} 2002)},
  Xseries       = {Lecture Notes in Computer Science},
  volume       = {2404},
  pages        = {359--364},
  publisher    = {Springer},
  year         = {2002},
  Xurl          = {https://doi.org/10.1007/3-540-45657-0\_29},
  doi          = {10.1007/3-540-45657-0_29},
  Xtimestamp    = {Thu, 14 Oct 2021 09:45:53 +0200},
  Xbiburl       = {https://dblp.org/rec/conf/cav/CimattiCGGPRST02.bib},
  Xbibsource    = {dblp computer science bibliography, https://dblp.org},
  address = {Copenhagen, Denmark}
}

@article{prasetiyowati2021determining,
  title={Determining threshold value on information gain feature selection to increase speed and prediction accuracy of random forest},
  author={Prasetiyowati, Maria Irmina and Maulidevi, Nur Ulfa and Surendro, Kridanto},
  journal={Journal of Big Data},
  volume={8},
  number={1},
  pages={84},
  year={2021},
  publisher={Springer},
  doi          = {10.1186/S40537-021-00472-4}
}

@article{hranisavljevic2020discretization,
  title={Discretization of hybrid CPPS data into timed automaton using restricted Boltzmann machines},
  author={Hranisavljevic, Nemanja and Maier, Alexander and Niggemann, Oliver},
  journal={Engineering Applications of Artificial Intelligence},
  volume={95},
  pages={103826},
  year={2020},
  publisher={Elsevier},
  doi = {10.1016/j.engappai.2020.103826}
}

@inproceedings{hranisavljevic2016novel,
  author    = {Nemanja Hranisavljevic and Oliver Niggemann and Alexander Maier},
  title     = {A Novel Anomaly Detection Algorithm for Hybrid Production Systems Based on Deep Learning and Timed Automata},
  booktitle = {Proceedings of the International Workshop on Principles of Diagnosis},
  year      = {2016},
  pages     = {1--8},
  doi ={10.48550/arXiv.2010.15415},
  address = {Denver, CO, USA},
  publisher = {International Workshop on the Principles of Diagnosis}
}

@inproceedings{medhat2015framework,
  title={A framework for mining hybrid automata from input/output traces},
  author={Medhat, Ramy and Ramesh, Sethu and Bonakdarpour, Borzoo and Fischmeister, Sebastian},
  booktitle={2015 International Conference on Embedded Software (EMSOFT)},
  pages={177--186},
  year={2015},
  publisher = {IEEE},
  address   = {Amsterdam, Netherlands},
  doi          = {10.1109/EMSOFT.2015.7318273}
}

@inproceedings{kampmann2020does,
  title={When does my program do this? learning circumstances of software behavior},
  author={Kampmann, Alexander and Havrikov, Nikolas and Soremekun, Ezekiel O and Zeller, Andreas},
  publisher = {Association for Computing Machinery},
  address   = {Virtual Event, USA},
  booktitle={Proceedings of the 28th ACM joint meeting on european software engineering conference and symposium on the foundations of software engineering},
  pages={1228--1239},
  year={2020},
  doi= {10.1145/3368089.3409687}
}

@article{clarke1986automatic,
  title={Automatic verification of finite-state concurrent systems using temporal logic specifications},
  author={Clarke, Edmund M and Emerson, E Allen and Sistla, A Prasad},
  journal={ACM Transactions on Programming Languages and Systems (TOPLAS)},
  volume={8},
  number={2},
  pages={244--263},
  year={1986},
  publisher={ACM New York, NY, USA},
  doi = {10.1145/5397.5399}
}

@InProceedings{gsm,
author="von Berg, Benjamin
and Aichernig, Bernhard K.",
Xeditor="Piskac, Ruzica
and Rakamari{\'{c}}, Zvonimir",
title="Extending AALpy with Passive Learning: A Generalized State-Merging Approach",
booktitle="Proceedings of 37th International Conference on Computer Aided Verification ({CAV} 2025)",
year="2025",
publisher="Springer Nature Switzerland",
address="Cham",
pages="127--140",
isbn="978-3-031-98685-7",
doi = "10.1007/978-3-031-98685-7_6"
}

@article{bartocci2020mining,
  title={Mining shape expressions from positive examples},
  author={Bartocci, Ezio and Deshmukh, Jyotirmoy and Gigler, Felix and Mateis, Cristinel and Ni{\v{c}}kovi{\'c}, Dejan and Qin, Xin},
  journal={IEEE Transactions on Computer-Aided Design of Integrated Circuits and Systems},
  volume={39},
  number={11},
  pages={3809--3820},
  year={2020},
  publisher={IEEE},
  doi = {10.1109/TCAD.2020.3012240}
}

@inproceedings{bartocci2021mining,
  title={Mining shape expressions with ShapeIt},
  author={Bartocci, Ezio and Deshmukh, Jyotirmoy and Mateis, Cristinel and Nesterini, Eleonora and Ni{\v{c}}kovi{\'c}, Dejan and Qin, Xin},
  booktitle={International Conference on Software Engineering and Formal Methods},
  pages={110--117},
  year={2021},
  publisher={Springer},
  doi ={10.1007/978-3-030-92124-8_7},
  address = {Virtual Event}
}

@inproceedings{agostinelli2021discovering,
  title={Discovering declarative process model behavior from event logs via model learning},
  author={Agostinelli, Simone and Bergami, Giacomo and Fiorenza, Alessio and Maggi, Fabrizio M and Marrella, Andrea and Patrizi, Fabio},
  booktitle={2021 3rd International Conference on Process Mining (ICPM)},
  pages={48--55},
  year={2021},
  publisher={IEEE},
  address = {Eindhoven, The Netherlands},
  doi          = {10.1109/ICPM53251.2021.9576870}
}

@book{linz2022introduction,
  title={An introduction to formal languages and automata},
  author={Linz, Peter and Rodger, Susan H},
  year={2022},
  publisher={Jones \& Bartlett Learning},
  address   = {Burlington, MA, USA}
}

@incollection{oncina1992identifying,
  title={Identifying regular languages in polynomial time},
  author={Oncina, Jos{\'e} and Garcia, Pedro},
  booktitle={Advances in structural and syntactic pattern recognition},
  pages={99--108},
  year={1992},
  publisher={World Scientific},
  address   = {Singapore},
  doi = {10.1142/9789812797919_0007}
}

@book{metaheuristicsbook, 
       author =    { Sean Luke }, 
       title =     { Essentials of Metaheuristics },
       edition =   { second },
       year =      { 2013 }, 
       publisher = { Lulu },
       address   = {Morrisville, NC, USA},
       note =      { Available for free at http://people.cs.gmu.edu/$\sim$sean/book/metaheuristics} 
     }

@article{ernst2007daikon,
  title={The Daikon system for dynamic detection of likely invariants},
  author={Ernst, Michael D and Perkins, Jeff H and Guo, Philip J and McCamant, Stephen and Pacheco, Carlos and Tschantz, Matthew S and Xiao, Chen},
  journal={Science of computer programming},
  volume={69},
  number={1-3},
  pages={35--45},
  year={2007},
  publisher={Elsevier},
  doi = {10.1016/j.scico.2007.01.015}
}
\end{document}